\documentclass[a4paper,11pt,showkeys]{revtex4-2}
\usepackage[utf8]{inputenc}
\usepackage[T1]{fontenc}
\usepackage{lmodern}
\usepackage{epsfig}
\usepackage{array}
\usepackage{amsmath}
\usepackage{longtable,tabularx,booktabs,comment}
\usepackage{epstopdf}
\usepackage{graphicx}
\graphicspath{/}

\begin{document}
	
\title{Search for Damped Oscillating Structures from Charged Pion Electromagnetic Form Factor Data}
\date{\today}
	
\medskip
	
\author{Erik Barto\v s}
\affiliation{Institute of Physics, Slovak Academy of Sciences, Dúbravská cesta 9,
	SK-84511 Bratislava, Slovak Republic}

\author{Stanislav Dubni\v cka}
\affiliation{Institute of Physics, Slovak Academy of Sciences, Dúbravská cesta 9,
	SK-84511 Bratislava, Slovak Republic}

\author{Anna Zuzana Dubni\v ckov\'a}
\affiliation{Department of Theoretical Physics, Comenius University, Mlynská dolina,
	SK-84248 Bratislava, Slovak Republic}

\begin{abstract}The damped oscillating structures recently revealed by a three parametric formula from the proton ``effective'' form factor data extracted of the measured total cross section $\sigma^{bare}_{tot}(e^+e^-\to p\bar p)$ still seem to have an unknown origin. The conjectures of their direct manifestation of the quark-gluon structure of the proton indicate that they are not specific only of the proton and neutron, but they have to be one's own, similar to other hadrons.
Therefore, the oscillatory structures from the charged pion electromagnetic form factor timelike data, extracted of the process $e^+e^-\to \pi^+ \pi^-$ are investigated by using the same procedure as in the case of the proton.
The analysis shows the appearance of the oscillating structures in the description of the charged pion electromagnetic form factor timelike data by three parametric formula with a rather large value of $\chi^2/ndf$, while the description of the data by the physically well-founded Unitary and Analytic model has not revealed any damped oscillating structures.
From the obtained result on the most simple object of strong interactions, one can conclude that damped oscillating structures received from the ``effective'' proton form factor data are probably generated by a utilization of the improper three parametric formula which does not describe these data with sufficient precision.
\end{abstract}


\maketitle

\section{Introduction}

The total cross section $\sigma^{bare}_{tot}(e^+e^- \to p \bar p)$ has been measured by two different methods, the~so called scan method \cite{Pedlar, Ablikim1, Ablikim2, Akhmetshin1, Akhmetshin2, Ablikim3} and the initial state radiation (ISR) technique~\cite{Aubert, Lees1, Lees2, Ablikim4, Ablikim5}.

The scan method consists of taking energy scan data on the process $e^+e^- \to p \bar p$, whereby the c.~m. energy of the $e^+e^-$ collider is systematically changed
from one energy value to another other~energy value.

The ISR technique provides data on the process $e^+e^- \to p \bar p \gamma$ at a fixed c.~m. energy value of the $e^+e^-$ collider at which maximal value of the luminosity is achieved, analysing events together with a photon emitted by the initial electron or positron, thus reducing the momentum transfer $q^2$ of the process. This method allows for the measurement of $\sigma^{bare}_{tot}(e^+e^- \to p \bar p)$ from the threshold of the reaction $e^+e^- \to p \bar p$ up to the fixed c.~m. energy \mbox{of the $e^+e^-$ collider}.

In practical applications of both methods for obtaining experimental information on $\sigma^{bare}_{tot}(e^+e^- \to p \bar p)$, and~so also on electromagnetic (EM) form factors (FFs), it was clearly demonstrated that by means of the ISR technique more precise information has been achieved. Therefore, in~this paper the data obtained by ISR technique will be~preferred.

A theoretical behavior of $\sigma^{bare}_{tot}(e^+e^- \to p \bar p)$ is described by the relation
\begin{eqnarray}\label{totcspp}
	\sigma_{tot}(e^+e^- \to p \bar p)=\frac{4 \pi \alpha^2 C_p \beta_p(s)}{3 s}
	\Big[|G^p_M(s)|^2+\frac{2m_p^2}{s}|G^p_E(s)|^2\Big],
\end{eqnarray}
with $\beta_p(s)=\sqrt{1-\frac{4 m^2_p}{s}}$, $\alpha$=1/137 and $C_p=\frac{\pi \alpha / \beta_p(s)}{1-\exp(-\pi \alpha / \beta_p(s))}$
to be the so-called Sommerfeld-Gamov-Sakharov Coulomb enhancement factor~\cite{BaPaZa}, which accounts for the EM interaction between the outgoing proton and~antiproton.

The functions $G^p_E(s)$ and $G^p_M(s)$ in Equation~(\ref{totcspp}) are the Sachs proton electric and proton magnetic FFs, respectively, depending on the c.~m. energy squared $s$ differently, however, at~the proton-antiproton threshold they are identical as it follows from their expressions through the proton Dirac $F_1^p(s)$ and the proton Pauli $F_2^p(s)$ FFs;
\begin{align}\label{EMFFbyDP}
	G^p_E(s)&= F_1^p(s) + \frac{s}{4m_p^2}F_2^p(s)\\
	G^p_M(s)&= F_1^p(s) + F_2^p(s).
\end{align}

As one could not determine both proton EM FFs from measured $\sigma^{bare}_{tot}(e^+e^- \to p \bar p)$ at any value $s>4m^2_p$ simultaneously, experimental groups in~\cite{Lees1, Ablikim2, Ablikim4, Ablikim3, Ablikim5}, with~the hope of achieving more information on the proton structure, generalized the threshold identity  $|G^p_E(4m^2_p)|\equiv|G^p_M(4m^2_p)|$ in Equation~(\ref{totcspp}) for all higher $s$-values up to $+\infty$, and~the data with errors obtained by means of the consequent expression
\begin{eqnarray}
	|G^p_{eff}(s)|=\sqrt{\frac{\sigma^{bare}_{tot}(e^+e^- \to p \bar p)}{\frac{4\pi\alpha^2C_p\beta_p(s)}{3s}\Big(1+\frac{2m^2_p}{s}\Big)}}
\end{eqnarray}
have been named the proton ``effective'' FF data. Immediately after the publishing of the BABAR Collaboration proton ``effective'' FF data~\cite{Lees1, Lees2}, the~modified form~\cite{TomRek} of the dipole formula for nucleon EM FFs behaviors in the spacelike region
\begin{eqnarray}\label{fortfunct}
	G^p_{eff}=\frac{A}{(1+s/m_a^2)(1-s/0.71\,\mathrm{GeV}^2)^2},
\end{eqnarray}
with the nucleon ``magic'' number $0.71$ GeV$^2$ has been applied successfully for their~description.

Here we would like to note, that the nucleon ``magic'' number $0.71$ GeV$^2$ in \mbox{Equation (\ref{fortfunct})} has its historical origin~\cite{DChCHRWW} in a simultaneous description of all spacelike data on the EM FFs of the proton $G^p_E(t)$, $G^p_M(t)$ obtained from the corresponding differential cross section by means of the Rosembluth method and the spacelike neutron EM FFs $G^n_E(t)$, $G^n_M(t)$ data by only one dipole formula
\begin{equation}\label{EMFFdip}
	G^p_E(t)\approx \frac{G^p_M(t)}{\mu_p}\approx  \frac{G^n_M(t)}{\mu_n}\approx -\frac{4m_n^2}{t}\frac{G^n_E(t)}{\mu_n}\approx \frac{1}{(1-\frac{t}{0.71\,\mathrm{GeV}^2})^2}
\end{equation}
where ``\emph{t}'' is the momentum transfer squared of the elastic scattering of electrons on nucleons and magnetic moments of the proton $\mu_p$ and the neutron $\mu_n$, respectively, which, however, is no longer valid for the following~reasons.

The differential cross section of an elastic scattering of electrons on protons looks as follows
\begin{equation}
	\frac{d \sigma}{d \Omega}= \left (\frac{d \sigma}{d \Omega}\right )_{Mott}\left\{\frac{\big(G^p_E(t)\big)^2 - \frac{t}{4 m_p^2} \big(G^p_M(t)\big)^2}{1 - \frac{t}{4 m_p^2}} - \frac{t}{2 m_p^2} \big(G^p_M(t)\big)^2 \tan^2 \theta/2 \right\}
\end{equation}
with $E$ the electron energy and $\theta$ the electron scattering angle,
\begin{equation}
	\left(\frac{d \sigma}{d \Omega}\right )_{Mott} = \frac{\alpha^2}{4E^2\sin^2\theta/2} \cdot\frac{\cos^2\theta/2}{1 + \frac{2E}{m_p}\sin^2\theta/2}
\end{equation}
represents the cross section for a structureless proton. Then, the ratio $\left(\frac{d \sigma}{d \Omega}\right )/\left(\frac{d \sigma}{d \Omega}\right )_{Mott}$ when evaluated for a fixed momentum transfer squared ``$-t'$' yields a straight line when plotted against $\tan^2 \theta/2$. The~slope of this straight line and intercept give the $G^p_E(t)$ and $G^p_M(t)$ FFs values. Such method of experimental determination of $G^p_E(t)$ indicated its dipole behaviour, but~with increased ``$-t$'' the reliability of determined $G^p_E(t)$ values acquired decreasing~tendency.

\begin{figure}[ht]
\includegraphics[width=0.65\textwidth]{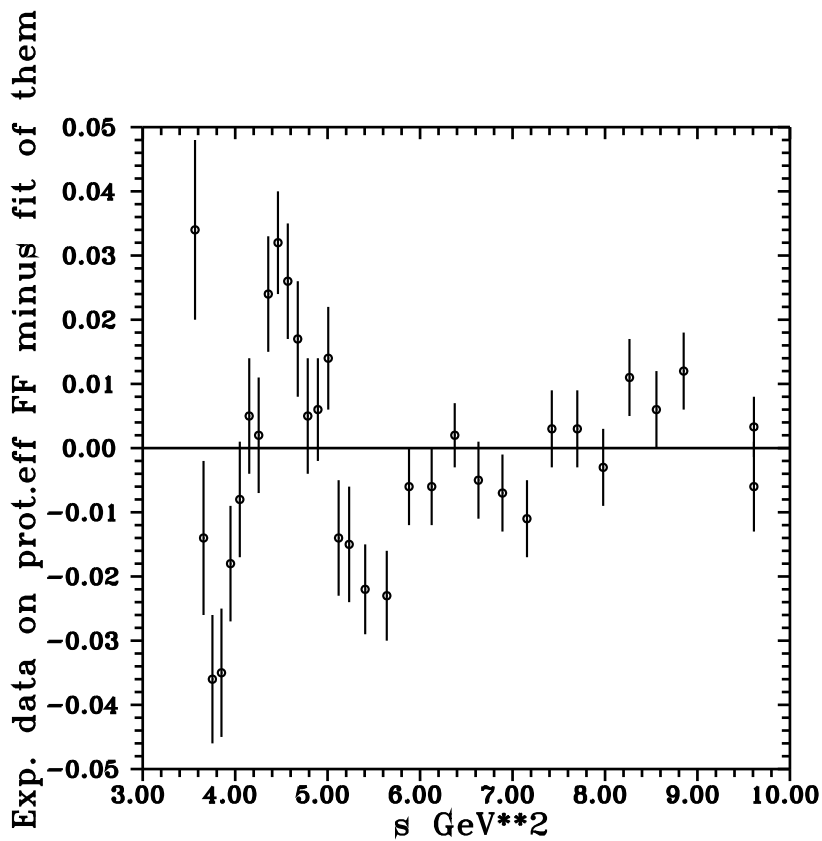}
\caption{Result of a 
 subtraction of the best fit of BABAR data~\cite{Lees1, Lees2} with the Formula (\ref{fortfunct}) from experimental values~\cite{Lees1, Lees2} with~errors.\label{fig:1}}
\end{figure}

Starting from 2000, the Akhiezer-Rekalo polarization method~\cite{AkhRek1,AkhRek2} consisting in the simultaneous measurements of the transverse
\begin{equation}
	P_t=\frac{h}{I_0}(-2)\sqrt{\tau(1+\tau)} G^N_EG^N_M \tan(\theta/2)
\end{equation}
and longitudinal
\begin{equation}
	P_l=\frac{h(E+E')}{I_0m_N}\sqrt{\tau(1+\tau)} G^{N2}_M \tan^2(\theta/2)
\end{equation}
components of the recoil nucleon's polarization in the electron scattering plane of the polarization transfer $\overrightarrow{e^-}N \to e^- \overrightarrow{N}$ process, with~$h$ as the electron beam helicity, $I_0$ as the unpolarized cross section excluding $\left (\frac{d \sigma}{d \Omega}\right )_{Mott}$ and $\tau=Q^2/{4m^2_N}=-s/4m^2_N$, new more reliable data in comparison with those obtained by means of the Rosenbluth method on the ratio $\frac{G^N_E(s)}{G^N_M(s)}=-\frac{P_t}{P_l}\frac{(E+E')}{2m_N}\tan(\theta/2)$ have been measured~\cite{Jones,Gayou,Punjabi,Zhan,Puckett2}, which clearly demonstrate that the electric proton FF in the spacelike region has no more dipole behavior \mbox{(it has a steeper fall)} and it can no more be described by the nucleon ``magic'' number in dipole formula \mbox{of Equation~(\ref{EMFFdip})}. Despite the latter, the best fit data given by \mbox{Equation (\ref{fortfunct})} and the value of the free parameters A = 7.7 and $m^2_a$ = 14.8 GeV$^2$ have been subtracted from BABAR data~\cite{Lees1, Lees2} on the ``effective'' FF with errors and in the plot of these differences with errors as a function of three momentum $p(s)=\sqrt{s(\frac{s}{4m^2_p}-1)}$ of one of the proton or antiproton in the frame where other one is at rest, damped oscillatory structures with regularly spaced maxima and minima have been revealed~\cite{BianTom} as presented in Figure~\ref{fig:1}.

The latter oscillatory behavior, however, is seen also if results of the previous analysis are plotted simply as a function of $\sqrt{s}$, (see Figure~13 of~\cite{Ablikim4}), consequently as a function of $s$ too, and~there is no need to plot them as a function of three momentum ``p(s)'' defined by the very special reference~frame.

Similar results are obtained in all other measurements of $\sigma^{bare}_{tot}(e^+e^-\to p\bar p)$ after the year 2013 and also in the recent measurement of $\sigma^{bare}_{tot}(e^+e^-\to n\bar n)$ \cite{Ablikim6}, however, in~this case with an opposite~behavior.

\begin{figure}[ht]
\includegraphics[width=0.8\textwidth]{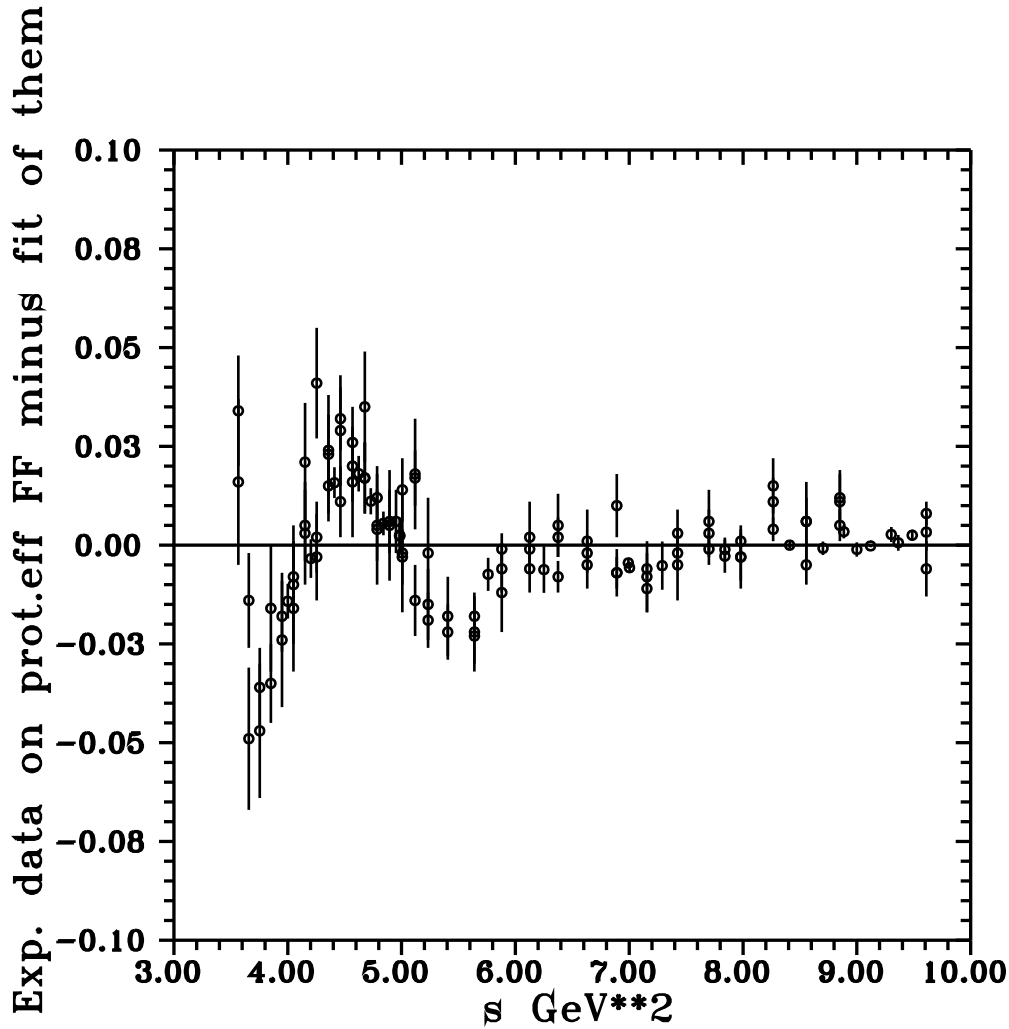}
\caption{Result of a subtraction of the best fit of all existing data on the proton ``effective'' form factor from these data with~errors.\label{fig:2}}
\end{figure}

In order to be sure of all these consequences, first we have collected all existing data on the proton ``effective'' FF measured until now from the papers~\cite{Lees1, Lees2, Ablikim2, Ablikim4, Ablikim3, Ablikim5} and then repeated simultaneous analysis of all of them by the Formula (\ref{fortfunct}). The~result with slightly different values of parameters $A=8.9\pm0.3$ and $m^2_a=9.2\pm0.8$ GeV$^2$ confirm the analysis of the paper~\cite{BianTom}, see Figure~\ref{fig:2},  however with $\chi^2/ndf\approx 5$ to be not acceptable from the statistical point of view declaring a good description of the experimental data. As~the value of $\chi^2/ndf$ is not quoted in the paper~\cite{BianTom}, we contend that also their value is not acceptable from the statistical point of view for a good description of analysed data too. On~the basis of the previous results, we arrive at a conclusion that it is likely an inaccurate description of data on the proton ``effective'' FF by Equation~(\ref{fortfunct}) is responsible for an appearance of the damped oscillatory structures from the proton ``effective'' FF~data.

If we are not right, the~origin of the damped oscillatory structures from the proton ``effective'' FF data~\cite{BianTom2} is still~unknown.

Some conjectures exist that the damped oscillatory structures are a direct manifestation of the quark-gluon structure of the proton~\cite{BianTom3}. If~so, then they can not be a specific only of the proton and neutron, and~they have to be one's own also of other~hadrons.

However, only in such hadrons investigations of an existence of oscillatory structures can be carried out for which a sufficiency of reliable EM FF data exists and there is also a physically well-founded theoretical model for their perfect~description.

So far, both requirements are entirely fulfilled by charged pion EM FF data and also by charged K-meson EM FF~data.

In this paper with an aim of a clarification of the latter problem, damped oscillatory structures from the data on the most simple strong interactions object, the~charged pion EM FF timelike data, are analysed in the framework of the model with a well-founded physical~background.

Attempts to explain the damped oscillations from the proton ``effective'' form factor data by means of a tangible physical background approach can be found \mbox{in the paper~\cite{Lorenz}}.

\section{Search for Damped Oscillation Structures in the Charged Pion EM FF Timelike~Data}

The timelike data on the charged pion EM FF $F^c_\pi(s)$ with errors are hidden in the measured total cross section $\sigma^{bare}_{tot}(e^+e^- \to \pi^+ \pi^-)$.

We would like to point out at once at the beginning that in the process of obtaining information on $|F^c_\pi(s)|$ from $\sigma^{bare}_{tot}(e^+e^- \to \pi^+ \pi^-)$, no nonphysical demands are needed, unlike the nucleon ``effective'' FF data, because~there is only one function $|F^c_\pi(s)|$ completely describing the measured total cross~section.

However, we meet another problem here. The~charged pion EM FF $F^c_\pi(s)$ represents the $\gamma \pi^+ \pi^-$ vertex generated by the strong interactions and not all of $\pi^+\pi^-$ pairs in the measured total cross section $\sigma^{bare}_{tot}(e^+e^- \to \pi^+ \pi^-)$ have strong interaction origin. Some part of them appears due to electromagnetic isospin violating decay of $\omega(782) \to \pi^+\pi^-$ responsible for a deformation of the right wing of the $\rho(770)$ meson peak, well known as the $\rho-\omega$-interference effect. Since we are investigating oscillation structures to be own of the electromagnetic FFs, one has to get rid of the electromagnetic isospin violating decay of $\omega(782) \to \pi^+\pi^-$ contribution somehow. Experimentalists are unable to isolate the electromagnetic isospin violating $\omega(782) \to \pi^+\pi^-$ decay contribution from~remaining.

In three of the most precise measurements until now~\cite{Lees3,Xiao,Ablikim7} of the total cross section $\sigma^{bare}_{tot}(e^+e^-\to\pi^+\pi^-(\gamma))$ with the initial state radiation (ISR) method, covering the energy range from the threshold up to 9 GeV$^2$, for~elimination of the electromagnetic isospin violating decay of $\omega(782) \to \pi^+\pi^-$ contribution the following procedure will be~applied.

First, the~total cross section of the $e^+e^- \to \pi^+\pi^-$ process is expressed by means of the absolute value squared of the sum of $F^c_\pi(s)$ and the isospin violating $\omega(782) \to \pi^+\pi^-$ decay contribution (further denoted by $F^{'}_\pi(s)$) in the form
\begin{equation}\label{totcspipi}
	\sigma^{bare}_{tot}(e^+e^- \to \pi^+ \pi^-)=\frac{\pi \alpha^2 \beta^3_\pi(s)}
	{3 s}{\Big|F^c_\pi(s)+Re^{i\phi}\frac{m^2_{\omega}}{m^2_{\omega}-s-im_{\omega}\Gamma_{\omega}}\Big|^2},
\end{equation}
where $F^c_\pi(s)$ is just the pure isovector charged pion EM FF to be expressed by the physically well founded Unitary and Analytic (U$\&$A) model given over the Formula (3.66) from~\cite{DD}
\begin{align}\label{pioFF}
	F^{th}_{\pi}[W(s)]=&(\frac{1-W^2}{1-W^2_N})^2\frac{(W-W_Z)(W_N-W_P)}{(W_N-W_Z)(W-W_P)}\nonumber\\
	&\times [\frac{(W_N-W_{\rho})(W_N-W^*_{\rho})(W_N-1/W_{\rho})(W_N-1/W^*_{\rho})}{(W-W_{\rho})(W-W^*_{\rho})
		(W-1/W_{\rho})(W-1/W^*_{\rho})}(\frac{f_{\rho\pi\pi}}{f_{\rho}})\\
	&+\sum_{v=\rho',\rho'',\rho'''}\frac{(W_N-W_v)(W_N-W^*_v)(W_N+W_v)(W_N+W^*_v)}
	{(W-W_v)(W-W^*_v)(W+W_v)(W+W^*_v)}(\frac{f_{v\pi\pi}}{f_v})],\nonumber
\end{align}
which respects all well known properties of the isovector EM FF of the charged pion, similar to the analyticity in the form of two square root type of branch points
approximation, first by the lowest threshold $s_0=4m^2_{\pi}$ and the second $s_{in}$ representing contributions of all higher eventual inelastic processes effectively, therefore it is a free parameter of the model, numerically evaluated in a fitting procedure of existing data.
\begin{equation}\label{comftrs}
	W(s)=i\frac{\sqrt{(\frac{s_{in}-s_0}{s_0})^{1/2}+(\frac{s-s_0}{s_0})^{1/2}}-
		\sqrt{(\frac{s_{in}-s_0}{s_0})^{1/2}-(\frac{s-s_0}{s_0})^{1/2}}}
	{\sqrt{(\frac{s_{in}-s_0}{s_0})^{1/2}+(\frac{s-s_0}{s_0})^{1/2}}+
		\sqrt{(\frac{s_{in}-s_0}{s_0})^{1/2}-(\frac{s-s_0}{s_0})^{1/2}}}
\end{equation}
is the conformal mapping of the four sheeted Riemann surface in $s$ variable into \mbox{one W-plane,}
\begin{equation}\label{norm}
	W_N=W(0)=i\frac{\sqrt{(\frac{s_{in}-s_0}{s_0})^{1/2}+i}-
		\sqrt{(\frac{s_{in}-s_0}{s_0})^{1/2}-i}}
	{\sqrt{(\frac{s_{in}-s_0}{s_0})^{1/2}+i}+
		\sqrt{(\frac{s_{in}-s_0}{s_0})^{1/2}-i}}
\end{equation}
is the normalization point in W-plane and
\begin{equation}\label{transVmes}
	W_v=W(s_v)=i\frac{\sqrt{(\frac{s_{in}-s_0}{s_0})^{1/2}+(\frac{s_v-s_0}{s_0})^{1/2}}-
		\sqrt{(\frac{s_{in}-s_0}{s_0})^{1/2}-(\frac{s_v-s_0}{s_0})^{1/2}}}
	{\sqrt{(\frac{s_{in}-s_0}{s_0})^{1/2}+(\frac{s_v-s_0}{s_0})^{1/2}}+
		\sqrt{(\frac{s_{in}-s_0}{s_0})^{1/2}-(\frac{s_v-s_0}{s_0})^{1/2}}}.
\end{equation}
is a position of poles corresponding to all isovector vector resonances forming the Model (\ref{pioFF}).

A normalization of the model to the electric charge leads to a reduction of the number of free coupling constant ratios $(f_{v\pi\pi}/{f_v})$ in Equation~(\ref{pioFF}) and the isovector nature of it is in the sense that only rho-meson and its excited states, i.e,. \mbox{$\rho(770), \rho'(1450), \rho''(1700)$ \cite{PDG}} and also the hypothetical $\rho'''(2150)$ \cite{BDEK}, in~order to cover the energetic region of data up to 9 GeV$^2$, can contribute to the FF behavior. The~reality condition $F^*_\pi(s)=F_\pi(s^*)$ has an effect of an appearance always two complex conjugate rho-meson poles on unphysical sheets. The~lefthand cut on the second Riemann sheet, revealed by the analytic continuation of the elastic FF unitarity condition, is approximated, in~a sense of Padé, by~one pole $W_P$ and one zero $W_Z$, with~their free~positions.

Further, in Equation~(\ref{totcspipi}) $\phi=\arctan\frac{m_{\omega}\Gamma_{\omega}}{m^2_{\rho}-m^2_{\omega}}$ is the $\rho-\omega$ interference phase and $R$ is the $\rho-\omega$ interference real~amplitude.

\begin{figure}[!ht]
	\hspace{-10pt}
	\includegraphics[width=\textwidth]{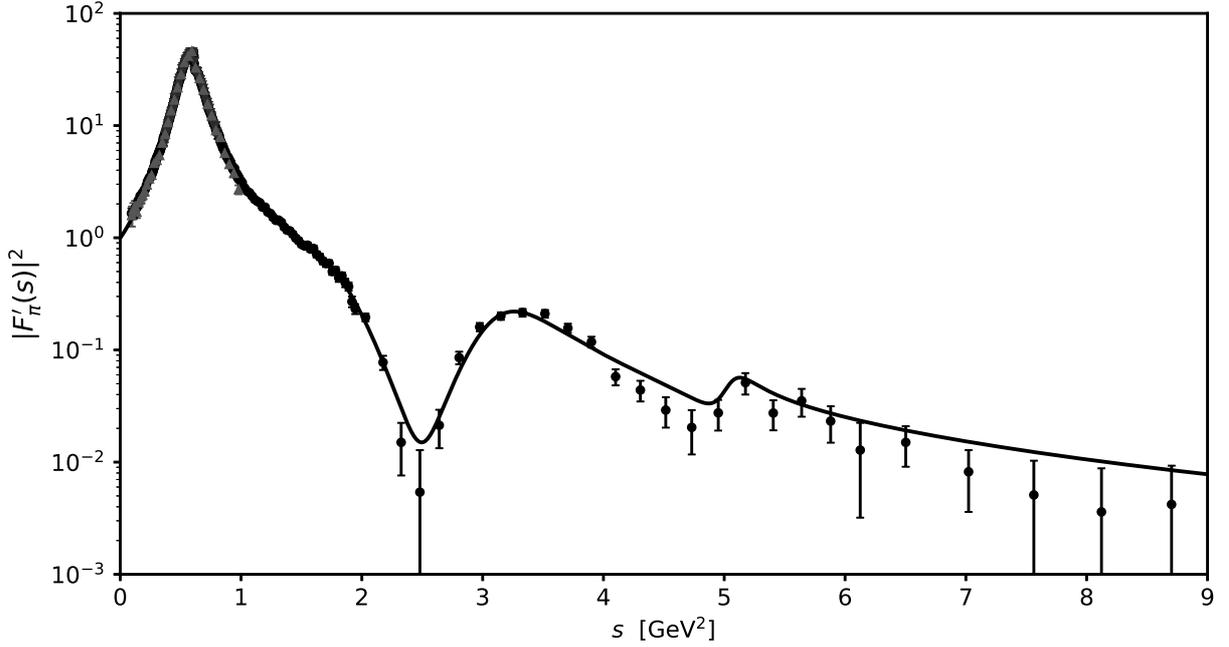}\\
	\caption{Optimal description of $|F^{'}_\pi(s)|^2$ data from~\cite{Lees3,Xiao,Ablikim7} in the energy range from the threshold {up to 9 GeV$^2$}.\label{fig:3}}
\end{figure}

\begin{figure}[!ht]
	\hspace{-10pt}
	\includegraphics[width=\textwidth]{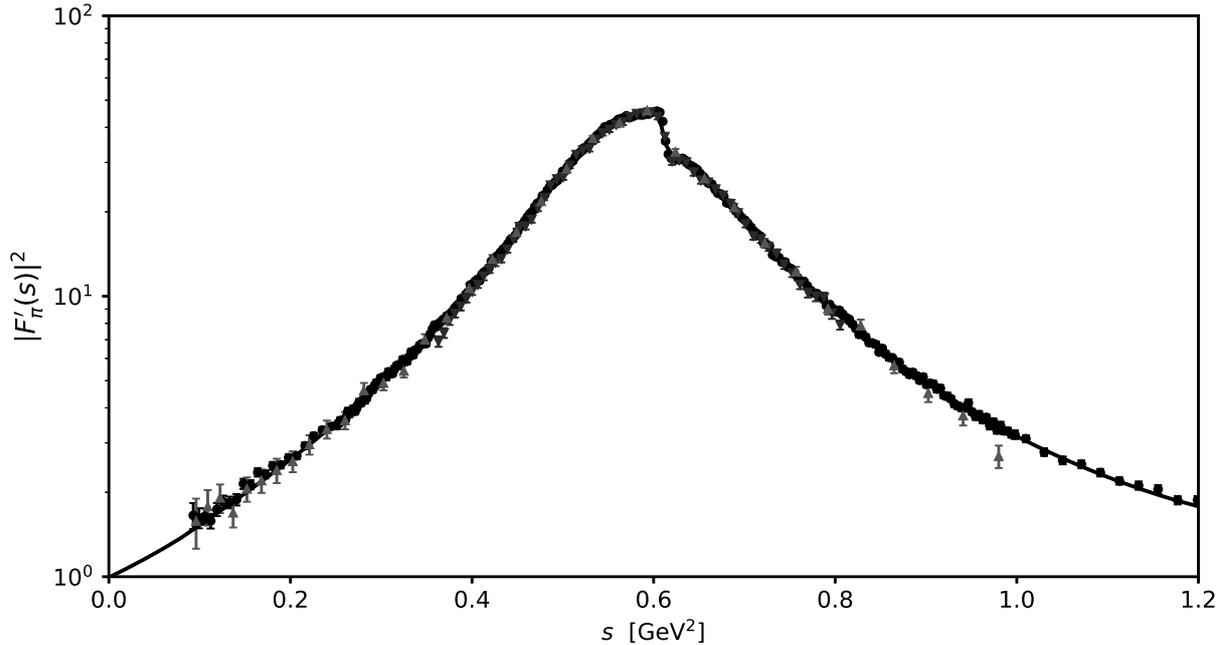}\\
	\caption{Optimal description of $|F^{'}_\pi(s)|^2$ data from~\cite{Lees3,Xiao,Ablikim7} in the region of the $\rho-\omega$-interference~effect.\label{fig:4}}
\end{figure}

In order to recognize all optimal parameter values of the model, an~analysis of existing data in~\cite{Lees3,Xiao,Ablikim7} on $|F^{'}_\pi(s)|^2$ has been carried out and the obtained results are presented numerically in Table~\ref{tab:1} and graphically by the curves in Figure~\ref{fig:3} and in detail of the $\rho-\omega$ interference effect in Figure~\ref{fig:4}.
\begin{table}[ht]
	\newcolumntype{C}{>{\centering\arraybackslash}X}
	\caption{Parameter values of the analysis of data in~\cite{Lees3,Xiao,Ablikim7} with minimum of $\chi^2/ndf=0.988$. \label{tab:1}}
	\begin{tabularx}{\textwidth}{cCC}
		\toprule
		$s_{in}=1.2730\pm0.0130$ (GeV$^2$) & $m_\rho =0.7620\pm0.0080$ (GeV)     & $\Gamma_\rho=0.1442\pm0.0014$ (GeV)  \\
		$(f_{\rho'\pi\pi}/f_{\rho'})=-0.0706 \pm 0.0012$ & $m_{\rho'}=1.3500\pm0.0110$ (GeV)  & $\Gamma_{\rho'}=0.3320\pm0.0033$ (GeV)\\
		$(f_{\rho''\pi\pi}/f_{\rho''})=0.0580 \pm 0.0010$ & $m_{\rho''}=1.7690\pm 0.0180$ (GeV) & $\Gamma_{\rho''}=0.2531\pm 0.0025$ (GeV)\\
		$(f_{\rho'''\pi\pi}/f_{\rho'''})=0.0021 \pm 0.0005$ & $m_{\rho'''}=2.2470\pm 0.0110$ (GeV) & $\Gamma_{\rho'''}=0.0700\pm 0.0007$ (GeV) \\
		$R=0.0113 \pm 0.0002$ & $W_Z=0.2845\pm 0.0033$ & $W_P=0.3830\pm 0.0060$ \\
		\bottomrule
	\end{tabularx}
\end{table}

Now, in~order to obtain the values of the pure isovector charged pion EM FF timelike data, the~absolute value squared relation in Equation (\ref{totcspipi}) is expressed as a product of the complex and the complex conjugate terms
\begin{multline}
		|F^c_\pi(s)+R e^{i\phi}\frac{m^2_{\omega}}{m^2_{\omega}-s-im_{\omega}\Gamma_{\omega}}|^2
		= \\ \{F^c_\pi(s)+R e^{i\phi}\frac{m^2_{\omega}}{m^2_{\omega}-s-im_{\omega}\Gamma_{\omega}}\}
		.\{F^{c*}_\pi(s)+R e^{-i\phi}\frac{m^2_{\omega}}{m^2_{\omega}-s+im_{\omega}\Gamma_{\omega}}\},
\end{multline}
and by using expressions $F^c_\pi(s) = |F^c_\pi(s)|e^{i\delta_\pi}$, $F^{c*}_\pi(s)=|F^c_\pi(s)|e^{-i\delta_\pi}$, also the identity of the pion EM FF phase with the P-wave isovector $\pi\pi$-phase shift $\delta_{\pi}(s)$ = $\delta^1_1(s)$, following from the charge pion EM FF elastic unitarity condition, practically considered to be valid up to 1~GeV$^2$, the~quadratic equation for the absolute value of the pure isovector charged pion EM FF $|F^c_\pi(s)|$ is found
\begin{eqnarray}
		|F^c_\pi(s)|^2+|F^c_\pi(s)| \frac{2 R m^2_{\omega}}{( m^2_{\omega}-s)^2+m^2_{\omega} \Gamma^2_{\omega}}
		[(m^2_{\omega}-s)\cos(\delta^1_1-\phi)+m_{\omega} \Gamma_{\omega}\sin(\delta^1_1-\phi)]\nonumber\\
		+\frac{R^2 m^4_{\omega}}{(m^2_{\omega}-s)^2+m^2_{\omega} \Gamma^2_{\omega}}-\frac{3s}{\pi \alpha^2 \beta^3_{\pi}(s)} \sigma^{bare}_{tot}(e^+e^- \to \pi^+ \pi^-) = 0.
\end{eqnarray}

Its solution gives the relation
\begin{align}\label{gensol}
		|F^c_\pi(s)|=&-\frac{ R m^2_{\omega}}{( m^2_{\omega}-s)^2+m^2_{\omega}\Gamma^2_{\omega}}
		[(m^2_{\omega}-s)\cos(\delta^1_1-\phi)+m_{\omega}\Gamma_{\omega}\sin(\delta^1_1-\phi)]\nonumber\\
		&\pm\{\frac{R^2 m^4_{\omega}}{[(m^2_{\omega}-s)^2+m^2_{\omega} \Gamma^2_{\omega}]^2}
		[(m^2_{\omega}-s)\cos(\delta^1_1-\phi)+m_{\omega}\Gamma_{\omega}\sin(\delta^1_1-\phi)]^2\\
		&-\frac{R^2 m^4_{\omega}}{(m^2_{\omega}-s)^2+m^2_{\omega} \Gamma^2_{\omega}}
		+\frac{3s}{\pi \alpha^2 \beta^3_{\pi}(s)} \sigma^{bare}_{tot}(e^+e^- \to \pi^+ \pi^-)\}^{1/2}\nonumber
\end{align}
in which a physical solution is given by the ``+'' sign of the second~term.

Then, by means of Equation (\ref{gensol}) with the ``+'' sign, the~most accurate up-to-now $\delta^1_1(s)$ data~\cite{Garcia} described by the most model independent parametrization~\cite{Bartos} with\mbox{ $q=[(s-4m^2_\pi)/4]^{1/2}$}, numerical values of parameters of Table~\ref{tab:1} and $\sigma^{bare}_{tot}(e^+e^- \to \pi^+ \pi^-)$ as measured in~\cite{Lees3}, the~data on the absolute value $|F^c_\pi(s)|$ of the pure isovector EM FF of the charged pion with errors are \mbox{evaluated in Table~\ref{tab:2}.}

\begin{longtable}{ccccccccc}
	\caption{The values of $|F^c_\pi(s)|$ with errors from measurements of J.N.~Lees~et~al.~\cite{Lees3}. \label{tab:2}}\\
	\hline
$s$ [GeV$^{2}$] & $|F_{\pi}^{c}(s)|$ & $s$ [GeV$^{2}$] & $|F_{\pi}^{c}(s)|$ & $s$ [GeV$^{2}$] & $|F_{\pi}^{c}(s)|$ & $s$ [GeV$^{2}$] & $|F_{\pi}^{c}(s)|$ \\
\hline
\endfirsthead
\caption{continued from previous page} \\
\endhead
$0.0930$ & $1.2877 \pm 0.0682$ & $0.3982$ & $3.3075 \pm 0.0278$ & $0.6384$ & $5.6773 \pm 0.0332$ & $0.9351$ & $2.0270 \pm 0.0272$ \\
$0.1056$ & $1.2825 \pm 0.0429$ & $0.4032$ & $3.3534 \pm 0.0282$ & $0.6448$ & $5.5609 \pm 0.0355$ & $0.9428$ & $2.0118 \pm 0.0307$ \\
$0.1190$ & $1.3199 \pm 0.0358$ & $0.4083$ & $3.3614 \pm 0.0282$ & $0.6512$ & $5.4027 \pm 0.0306$ & $0.9506$ & $1.9786 \pm 0.0233$ \\
$0.1332$ & $1.3583 \pm 0.0328$ & $0.4134$ & $3.4846 \pm 0.0292$ & $0.6577$ & $5.2312 \pm 0.0342$ & $0.9584$ & $1.9563 \pm 0.0286$ \\
$0.1482$ & $1.4659 \pm 0.0309$ & $0.4186$ & $3.5426 \pm 0.0289$ & $0.6642$ & $5.1332 \pm 0.0335$ & $0.9663$ & $1.9340 \pm 0.0250$ \\
$0.1640$ & $1.5359 \pm 0.0266$ & $0.4238$ & $3.6764 \pm 0.0266$ & $0.6708$ & $4.9312 \pm 0.0304$ & $0.9742$ & $1.8966 \pm 0.0279$ \\
$0.1806$ & $1.5775 \pm 0.0258$ & $0.4290$ & $3.7338 \pm 0.0292$ & $0.6773$ & $4.8596 \pm 0.0331$ & $0.9821$ & $1.8682 \pm 0.0310$ \\
$0.1980$ & $1.6306 \pm 0.0244$ & $0.4343$ & $3.8258 \pm 0.0295$ & $0.6839$ & $4.7191 \pm 0.0322$ & $0.9900$ & $1.8280 \pm 0.0253$ \\
$0.2162$ & $1.7123 \pm 0.0232$ & $0.4396$ & $3.9133 \pm 0.0283$ & $0.6906$ & $4.5615 \pm 0.0322$ & $0.9980$ & $1.8009 \pm 0.0314$ \\
$0.2352$ & $1.8263 \pm 0.0235$ & $0.4449$ & $3.9924 \pm 0.0271$ & $0.6972$ & $4.4369 \pm 0.0315$ & $1.0302$ & $1.6764 \pm 0.0263$ \\
$0.2510$ & $1.8588 \pm 0.0271$ & $0.4502$ & $4.1074 \pm 0.0293$ & $0.7039$ & $4.3270 \pm 0.0284$ & $1.0712$ & $1.5936 \pm 0.0257$ \\
$0.2550$ & $1.8997 \pm 0.0243$ & $0.4556$ & $4.2332 \pm 0.0306$ & $0.7106$ & $4.1883 \pm 0.0319$ & $1.1130$ & $1.4881 \pm 0.0225$ \\
$0.2591$ & $1.9009 \pm 0.0288$ & $0.4610$ & $4.3613 \pm 0.0294$ & $0.7174$ & $4.1039 \pm 0.0318$ & $1.1556$ & $1.4378 \pm 0.0252$ \\
$0.2632$ & $1.9705 \pm 0.0300$ & $0.4665$ & $4.4600 \pm 0.0308$ & $0.7242$ & $3.9646 \pm 0.0298$ & $1.1990$ & $1.3746 \pm 0.0230$ \\
$0.2673$ & $1.9726 \pm 0.0239$ & $0.4720$ & $4.6090 \pm 0.0285$ & $0.7310$ & $3.7943 \pm 0.0308$ & $1.2432$ & $1.2880 \pm 0.0247$ \\
$0.2714$ & $1.9720 \pm 0.0264$ & $0.4775$ & $4.7514 \pm 0.0320$ & $0.7379$ & $3.7528 \pm 0.0274$ & $1.2882$ & $1.1999 \pm 0.0228$ \\
$0.2756$ & $2.0446 \pm 0.0304$ & $0.4830$ & $4.8630 \pm 0.0289$ & $0.7448$ & $3.6812 \pm 0.0309$ & $1.3340$ & $1.1756 \pm 0.0247$ \\
$0.2798$ & $2.0663 \pm 0.0241$ & $0.4886$ & $4.9630 \pm 0.0312$ & $0.7517$ & $3.5788 \pm 0.0265$ & $1.3806$ & $1.0842 \pm 0.0227$ \\
$0.2841$ & $2.0875 \pm 0.0300$ & $0.4942$ & $5.0519 \pm 0.0326$ & $0.7586$ & $3.4504 \pm 0.0308$ & $1.4280$ & $1.0412 \pm 0.0242$ \\
$0.2884$ & $2.1566 \pm 0.0243$ & $0.4998$ & $5.2457 \pm 0.0329$ & $0.7656$ & $3.3948 \pm 0.0265$ & $1.4762$ & $0.9737 \pm 0.0249$ \\
$0.2927$ & $2.1858 \pm 0.0298$ & $0.5055$ & $5.3471 \pm 0.0333$ & $0.7726$ & $3.2740 \pm 0.0308$ & $1.5252$ & $0.9220 \pm 0.0231$ \\
$0.2970$ & $2.2061 \pm 0.0283$ & $0.5112$ & $5.4675 \pm 0.0341$ & $0.7797$ & $3.2273 \pm 0.0307$ & $1.5750$ & $0.8929 \pm 0.0239$ \\
$0.3014$ & $2.2659 \pm 0.0253$ & $0.5170$ & $5.6229 \pm 0.0318$ & $0.7868$ & $3.1704 \pm 0.0263$ & $1.6256$ & $0.8460 \pm 0.0255$ \\
$0.3058$ & $2.2700 \pm 0.0296$ & $0.5227$ & $5.7551 \pm 0.0351$ & $0.7939$ & $3.0846 \pm 0.0295$ & $1.6770$ & $0.7899 \pm 0.0250$ \\
$0.3102$ & $2.3078 \pm 0.0301$ & $0.5285$ & $5.8519 \pm 0.0313$ & $0.8010$ & $2.9917 \pm 0.0275$ & $1.7292$ & $0.7714 \pm 0.0249$ \\
$0.3147$ & $2.3458 \pm 0.0284$ & $0.5344$ & $5.9386 \pm 0.0334$ & $0.8082$ & $2.9634 \pm 0.0256$ & $1.7822$ & $0.7180 \pm 0.0250$ \\
$0.3192$ & $2.3847 \pm 0.0298$ & $0.5402$ & $6.0875 \pm 0.0359$ & $0.8154$ & $2.9001 \pm 0.0308$ & $1.8360$ & $0.6780 \pm 0.0248$ \\
$0.3238$ & $2.4373 \pm 0.0291$ & $0.5461$ & $6.2520 \pm 0.0368$ & $0.8226$ & $2.7916 \pm 0.0253$ & $1.8906$ & $0.6083 \pm 0.0252$ \\
$0.3283$ & $2.4188 \pm 0.0301$ & $0.5520$ & $6.2996 \pm 0.0369$ & $0.8299$ & $2.7271 \pm 0.0287$ & $1.9460$ & $0.4836 \pm 0.0242$ \\
$0.3329$ & $2.5142 \pm 0.0300$ & $0.5580$ & $6.3527 \pm 0.0374$ & $0.8372$ & $2.6327 \pm 0.0278$ & $2.1756$ & $0.2798 \pm 0.0201$ \\
$0.3376$ & $2.5335 \pm 0.0303$ & $0.5640$ & $6.4270 \pm 0.0343$ & $0.8446$ & $2.6144 \pm 0.0311$ & $2.4806$ & $0.0746 \pm 0.0504$ \\
$0.3422$ & $2.5808 \pm 0.0276$ & $0.5700$ & $6.4872 \pm 0.0361$ & $0.8519$ & $2.5705 \pm 0.0300$ & $2.8056$ & $0.2934 \pm 0.0186$ \\
$0.3469$ & $2.6007 \pm 0.0291$ & $0.5761$ & $6.4301 \pm 0.0356$ & $0.8593$ & $2.4773 \pm 0.0305$ & $3.1506$ & $0.4488 \pm 0.0172$ \\
$0.3516$ & $2.6623 \pm 0.0247$ & $0.5822$ & $6.4563 \pm 0.0360$ & $0.8668$ & $2.4332 \pm 0.0299$ & $3.5156$ & $0.4595 \pm 0.0183$ \\
$0.3564$ & $2.7621 \pm 0.0295$ & $0.5883$ & $6.3945 \pm 0.0360$ & $0.8742$ & $2.3619 \pm 0.0294$ & $3.9006$ & $0.3440 \pm 0.0191$ \\
$0.3612$ & $2.7882 \pm 0.0263$ & $0.5944$ & $6.3437 \pm 0.0408$ & $0.8817$ & $2.3077 \pm 0.0260$ & $4.3056$ & $0.2100 \pm 0.0220$ \\
$0.3660$ & $2.8463 \pm 0.0252$ & $0.6006$ & $6.2978 \pm 0.0423$ & $0.8892$ & $2.2963 \pm 0.0283$ & $4.7306$ & $0.1433 \pm 0.0305$ \\
$0.3709$ & $2.8903 \pm 0.0285$ & $0.6068$ & $6.1805 \pm 0.0393$ & $0.8968$ & $2.2847 \pm 0.0275$ & $5.1756$ & $0.2265 \pm 0.0246$ \\
$0.3758$ & $2.9375 \pm 0.0224$ & $0.6131$ & $5.9525 \pm 0.0367$ & $0.9044$ & $2.2257 \pm 0.0297$ & $5.6406$ & $0.1880 \pm 0.0261$ \\
$0.3807$ & $3.0037 \pm 0.0296$ & $0.6194$ & $6.0538 \pm 0.0391$ & $0.9120$ & $2.1731 \pm 0.0295$ & $6.1256$ & $0.1135 \pm 0.0424$ \\
$0.3856$ & $3.0602 \pm 0.0277$ & $0.6257$ & $5.9463 \pm 0.0339$ & $0.9197$ & $2.1173 \pm 0.0257$ & $7.0225$ & $0.0908 \pm 0.0254$ \\
$0.3906$ & $3.1212 \pm 0.0273$ & $0.6320$ & $5.8797 \pm 0.0374$ & $0.9274$ & $2.0884 \pm 0.0241$ & $8.1225$ & $0.0602 \pm 0.0433$ \\
$0.3956$ & $3.2098 \pm 0.0265$ & --- & --- & --- & --- & --- & --- \\
\hline
\end{longtable}
$\sigma^{bare}_{tot}(e^+e^- \to \pi^+ \pi^-)$ as measured in~\cite{Xiao} the data on the absolute value $|F^c_\pi(s)|$ of the pure isovector EM FF of the charged pion with errors are evaluated in Table~\ref{tab:3}, and $\sigma^{bare}_{tot}(e^+e^- \to \pi^+ \pi^-)$ as measured in~\cite{Ablikim7} the data on the absolute value $|F^c_\pi(s)|$ of the pure isovector EM FF of the charged pion with errors are evaluated in Table~\ref{tab:4}.
\begin{longtable}{ccccccccc}
	\caption{The values of $|F^c_\pi(s)|$ with errors from measurements of T.~Xiao~et~al.~\cite{Xiao}. \label{tab:3}}\\
	\hline
$s$ [GeV$^{2}$] & $|F_{\pi}^{c}(s)|$ & $s$ [GeV$^{2}$] & $|F_{\pi}^{c}(s)|$ & $s$ [GeV$^{2}$] & $|F_{\pi}^{c}(s)|$ & $s$ [GeV$^{2}$] & $|F_{\pi}^{c}(s)|$ \\
\hline
\endfirsthead
\caption{continued from previous page} \\
\endhead
$0.0961$ & $1.2579 \pm 0.1274$ & $0.2401$ & $1.8320 \pm 0.0678$ & $0.4489$ & $4.0967 \pm 0.0645$ & $0.7225$ & $3.9831 \pm 0.0712$  \\
$0.1089$ & $1.3352 \pm 0.0964$ & $0.2601$ & $1.8998 \pm 0.0676$ & $0.4761$ & $4.6435 \pm 0.0666$ & $0.7569$ & $3.5322 \pm 0.0712$  \\
$0.1225$ & $1.3827 \pm 0.0809$ & $0.2809$ & $2.1465 \pm 0.0695$ & $0.5041$ & $5.2939 \pm 0.0657$ & $0.7921$ & $3.0313 \pm 0.0717$  \\
$0.1369$ & $1.3011 \pm 0.0739$ & $0.3025$ & $2.2129 \pm 0.0638$ & $0.5329$ & $5.9774 \pm 0.0666$ & $0.8281$ & $2.8224 \pm 0.0731$  \\
$0.1521$ & $1.4347 \pm 0.0712$ & $0.3249$ & $2.3250 \pm 0.0614$ & $0.5625$ & $6.3216 \pm 0.0681$ & $0.8649$ & $2.3964 \pm 0.0730$  \\
$0.1681$ & $1.4848 \pm 0.0713$ & $0.3481$ & $2.6404 \pm 0.0637$ & $0.5929$ & $6.4618 \pm 0.0715$ & $0.9025$ & $2.1373 \pm 0.0742$  \\
$0.1849$ & $1.5489 \pm 0.0772$ & $0.3721$ & $2.8870 \pm 0.0631$ & $0.6241$ & $6.1518 \pm 0.0733$ & $0.9409$ & $1.9482 \pm 0.0749$  \\
$0.2025$ & $1.6065 \pm 0.0683$ & $0.3969$ & $3.2578 \pm 0.0644$ & $0.6561$ & $5.2718 \pm 0.0702$ & $0.9801$ & $1.6492 \pm 0.0752$  \\
$0.2209$ & $1.7209 \pm 0.0685$ & $0.4225$ & $3.6715 \pm 0.0648$ & $0.6889$ & $4.6328 \pm 0.0705$ & --- & --- \\
\hline
\end{longtable}

\begin{longtable}{ccccccccc}
	\caption{The values of $|F^c_\pi(s)|$ with errors from measurements of M.~Ablikim~et~al.~\cite{Ablikim7}.\label{tab:4}}\\
	\hline
$s$ [GeV$^{2}$] & $|F_{\pi}^{c}(s)|$ & $s$ [GeV$^{2}$] & $|F_{\pi}^{c}(s)|$ & $s$ [GeV$^{2}$] & $|F_{\pi}^{c}(s)|$ & $s$ [GeV$^{2}$] & $|F_{\pi}^{c}(s)|$ \\
\hline
\endfirsthead
\caption{continued from previous page} \\
\endhead
$0.3630$ &  $2.6228 \pm 0.0571$ & $0.4590$ &  $4.2006 \pm 0.0593$ &  $0.5663$ & $6.3316 \pm 0.0773$ & $0.6848$  & $4.7102 \pm 0.0649$ \\
$0.3691$ &  $2.7159 \pm 0.0551$ & $0.4658$ &  $4.3154 \pm 0.0577$ &  $0.5738$ & $6.4403 \pm 0.0757$ & $0.6931$  & $4.5238 \pm 0.0674$ \\
$0.3752$ &  $2.8587 \pm 0.0524$ & $0.4727$ &  $4.5158 \pm 0.0551$ &  $0.5814$ & $6.4891 \pm 0.0747$ & $0.7014$  & $4.3203 \pm 0.0588$ \\
$0.3813$ &  $2.9442 \pm 0.0509$ & $0.4796$ &  $4.7177 \pm 0.0527$ &  $0.5891$ & $6.4405 \pm 0.0746$ & $0.7098$  & $4.1088 \pm 0.0617$ \\
$0.3875$ &  $3.0272 \pm 0.0495$ & $0.4865$ &  $4.9612 \pm 0.0601$ &  $0.5968$ & $6.3509 \pm 0.0744$ & $0.7183$  & $4.0533 \pm 0.0625$ \\
$0.3938$ &  $3.1239 \pm 0.0479$ & $0.4935$ &  $5.0862 \pm 0.0586$ &  $0.6045$ & $6.0821 \pm 0.0758$ & $0.7268$  & $3.9213 \pm 0.0646$ \\
$0.4001$ &  $3.2176 \pm 0.0465$ & $0.5006$ &  $5.1210 \pm 0.0582$ &  $0.6123$ & $5.9440 \pm 0.0834$ & $0.7353$  & $3.8124 \pm 0.0663$ \\
$0.4064$ &  $3.3085 \pm 0.0452$ & $0.5077$ &  $5.3716 \pm 0.0647$ &  $0.6202$ & $5.9887 \pm 0.0821$ & $0.7439$  & $3.6460 \pm 0.0693$ \\
$0.4128$ &  $3.4262 \pm 0.0582$ & $0.5148$ &  $5.6100 \pm 0.0619$ &  $0.6281$ & $5.8975 \pm 0.0724$ & $0.7526$  & $3.5157 \pm 0.0575$ \\
$0.4193$ &  $3.5256 \pm 0.0566$ & $0.5220$ &  $5.7086 \pm 0.0694$ &  $0.6360$ & $5.7515 \pm 0.0729$ & $0.7613$  & $3.3510 \pm 0.0603$ \\
$0.4258$ &  $3.6222 \pm 0.0551$ & $0.5293$ &  $5.7354 \pm 0.0690$ &  $0.6440$  & $5.4657 \pm 0.0760$ & $0.7700$  & $3.2412 \pm 0.0623$ \\
$0.4323$ &  $3.6757 \pm 0.0542$ & $0.5366$ &  $5.9464 \pm 0.0582$ &  $0.6521$  & $5.2508 \pm 0.0688$ & $0.7788$  & $3.1919 \pm 0.0632$ \\
$0.4389$ &  $3.8206 \pm 0.0522$ & $0.5439$ &  $6.0994 \pm 0.0647$ &  $0.6602$  & $5.1830 \pm 0.0693$ & $0.7877$  & $3.1741 \pm 0.0636$ \\
$0.4456$ &  $3.9851 \pm 0.0500$ & $0.5513$ &  $6.1742 \pm 0.0638$ &  $0.6683$  & $5.0217 \pm 0.0713$ & $0.7966$  & $2.9754 \pm 0.0678$ \\
$0.4523$ &  $4.1906 \pm 0.0594$ & $0.5588$ &  $6.2835 \pm 0.0704$ &  $0.6765$  & $4.8821 \pm 0.0627$ & $0.8055$  & $2.8349 \pm 0.0534$ \\
\hline
\end{longtable}

\begin{figure}[ht]
	
	\includegraphics[width=.5\textwidth]{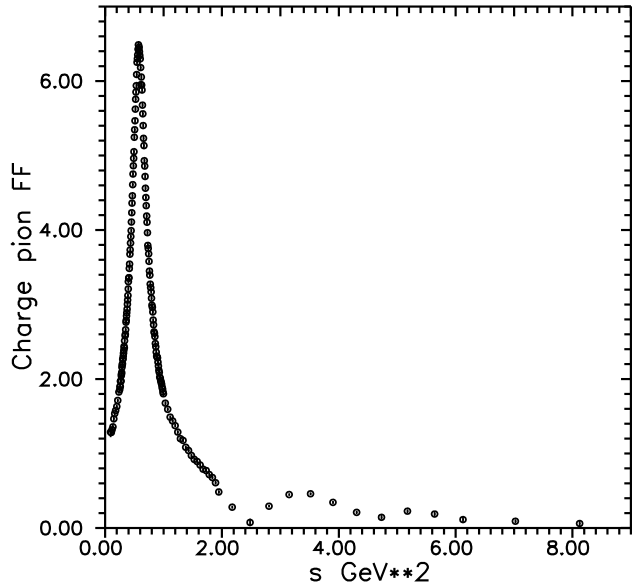}\\
	\caption{Calculated values of $|F^c_{\pi}(s)|$ with errors from measured $\sigma^{bare}_{tot}(e^+e^- \to \pi^+ \pi^-)$ in~\cite{Lees3,Xiao,Ablikim7} by the Expression (\ref{gensol}).\label{fig:5}}
\end{figure}

All three sets of data on the absolute value $|F^c_{\pi}(s)|$ of the pure isovector EM FF of the charged pion as a function of $s$ of threshold up to 9 GeV$^2$ as given in Tables~\ref{tab:2}--\ref{tab:4} are graphically presented in Figure~\ref{fig:5}.

Afterwards these data are optimally described by a similar formula to Equation (\ref{fortfunct}), however, now the nucleon ``magic'' number 0.71 is substituted by the third free parameter $A_3$, because~one can not expect to be suitable nucleon ``magic'' number for optimal description of the absolute value $|F^c_{\pi}(s)|$ data and one has to look for the charged pion ``magic'' number in its~place.

\begin{figure}[ht]
	
	\includegraphics[width=0.55\textwidth]{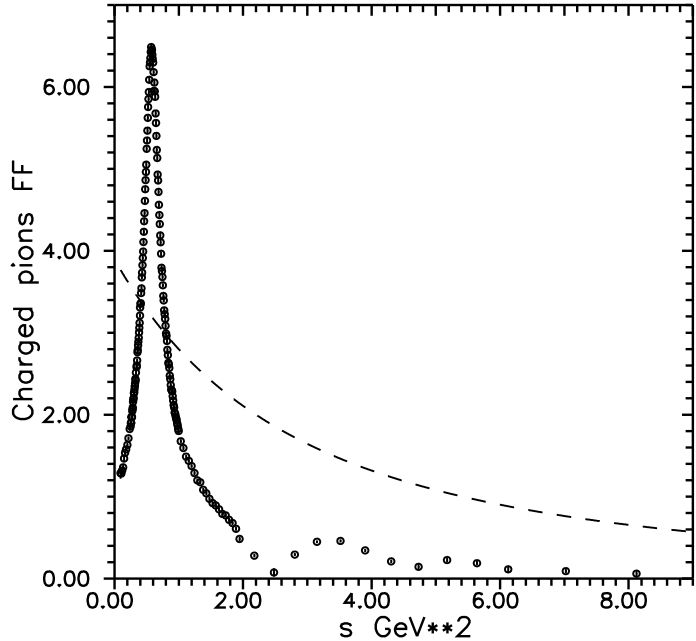}
	\caption{Optimal description of $|F^c_{\pi}(s)|$ data with dashed line given by a similar formula \mbox{to Equation (\ref{fortfunct})}, however now with the third parameter to be $A_3=-5.5647\pm0.1037 $ GeV$^{2}$. \label{fig:6}}
\end{figure}

The best fit of the data presented in Figure~\ref{fig:5} has been achieved with parameter values ${A=3.9888\pm0.0061}$, ${m^2_a=5.5647\pm0.1915 }$  GeV$^{2}$ and the charged pion ``magic'' number $A_3=$ $-5.5647\pm0.1037$ GeV$^{2}$. The~result of the fit is graphically presented in Figure~\ref{fig:6} by the dashed~line.

\begin{figure}[ht]
	
	\includegraphics[width=.5\textwidth]{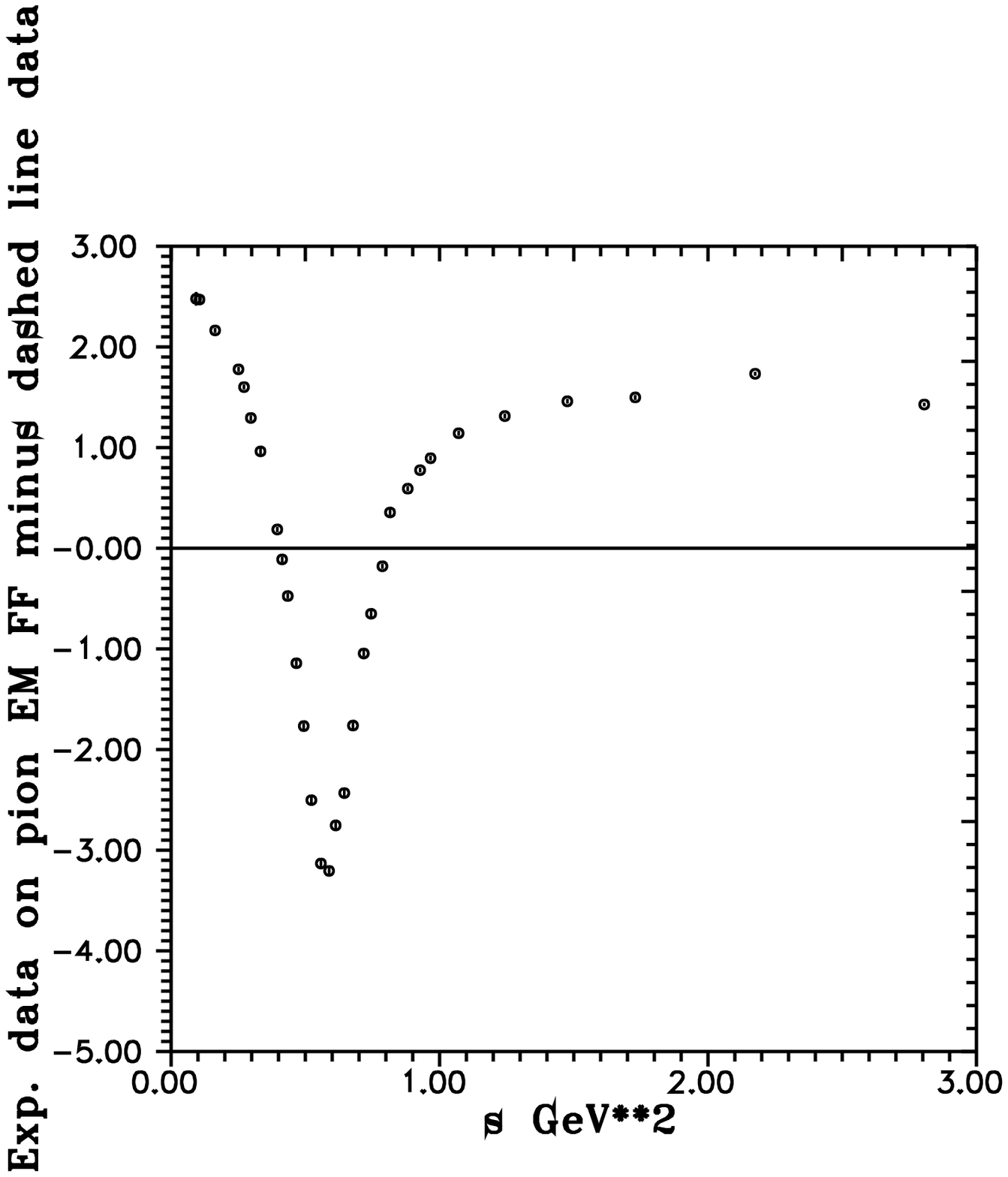}
	\caption{Damped oscillation structures obtained by a subtraction of dashed line data in Figure~\ref{fig:6} from exp. data on $|F^c_{\pi}(s)|$ in Tables~\ref{fig:2}--\ref{fig:4}. \label{fig:7}}
\end{figure}

If dashed line values are subtracted from the experimental values of $|F^c_{\pi}(s)|$ with errors in Tables~\ref{fig:2}--\ref{fig:4}, damped oscillating structures from charged pion electromagnetic form factor timelike data appear, as~it is presented in Figure~\ref{fig:7}.

\begin{figure}[ht]
	
	\includegraphics[width=.65\columnwidth]{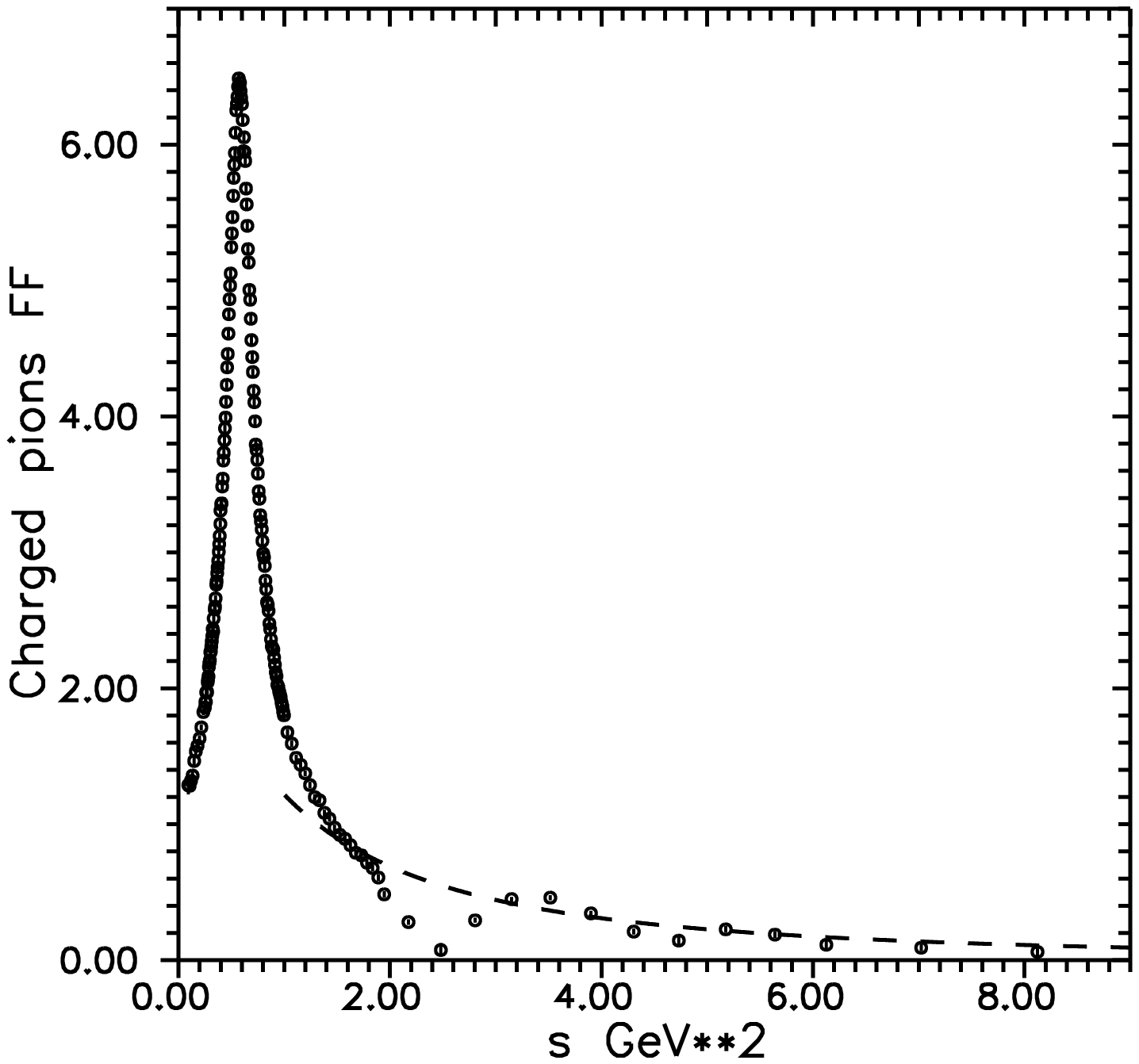}
	\caption{Optimal description of $|F^c_{\pi}(s)|$ data with dashed line given by a similar formula \mbox{to Equation~(\ref{fortfunct})}, however now with the third parameter to be $A_3=-2.0369\pm0.2613$ GeV$^2$. \label{fig:8}}
\end{figure}

\begin{figure}[ht]
	
	\includegraphics[width=.65\columnwidth]{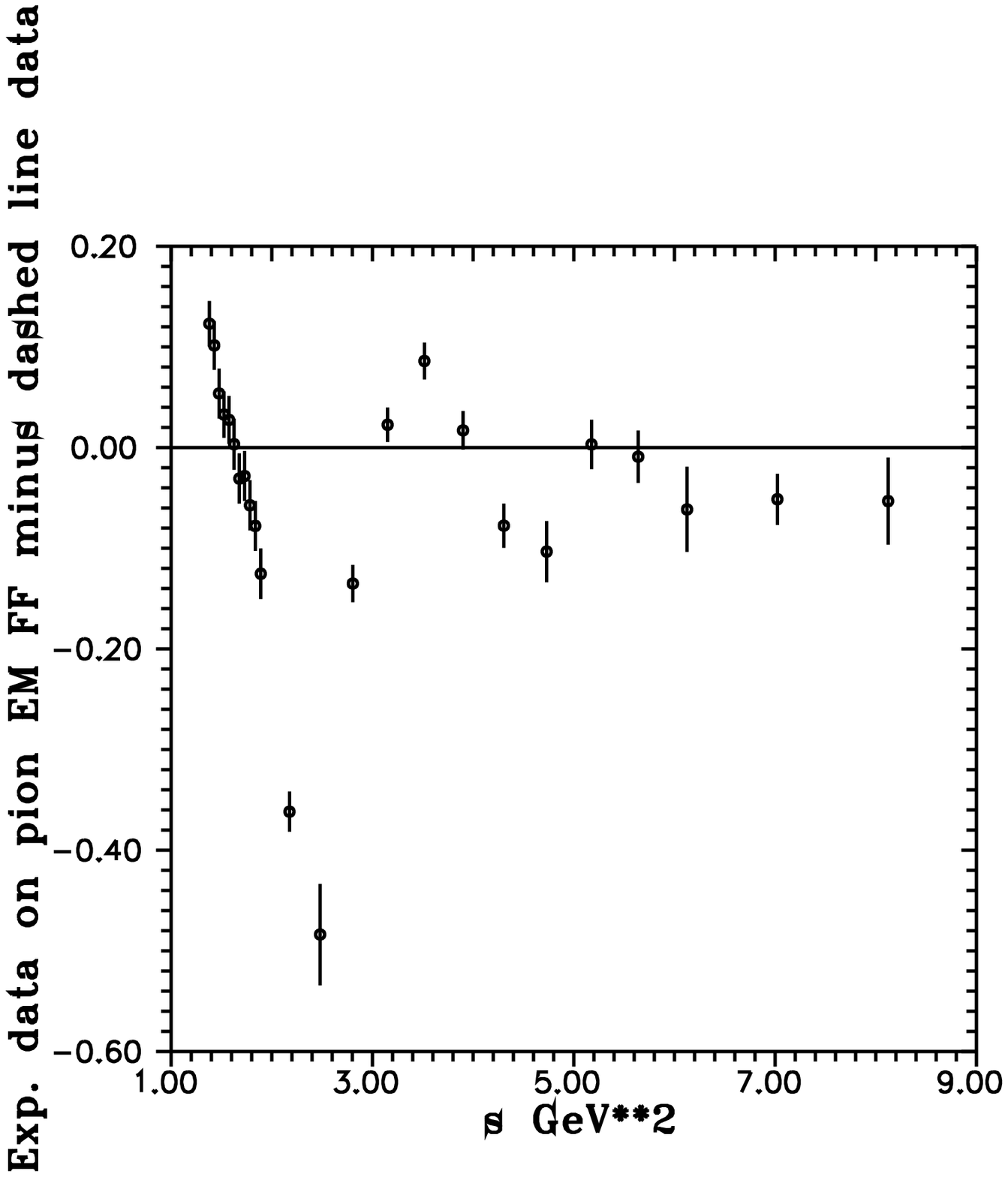}
	\caption{Damped oscillation structures obtained by a subtraction of dashed line data in Figure~\ref{fig:8} from exp. data on $|F^c_{\pi}(s)|$ in Tables~\ref{fig:2}--\ref{fig:4}. \label{fig:9}}
\end{figure}

If only the data on $|F^c_{\pi}(s)|$ from Figure~\ref{fig:5} just behind the distinct $\rho$-meson peak, starting from \mbox{1 GeV$^2$} up to 9 GeV$^2$, are optimally described by a
similar formula to Equation (\ref{fortfunct}) with all three free parameters with values ${A=3.7485\pm0.4463}$, ${m^2_a=3.9681\pm0.6064}$ GeV$^{2}$, \mbox{${A_3=-2.0369\pm0.2613}$ GeV$^{2}$}, one obtains the result graphically presented by dashed line in\mbox{ Figure~\ref{fig:8}},
and the corresponding damped oscillating structures appear as it is\mbox{ presented in Figure~\ref{fig:9}.}

\unskip

\begin{figure}[ht]
	
	\includegraphics[width=.65\columnwidth]{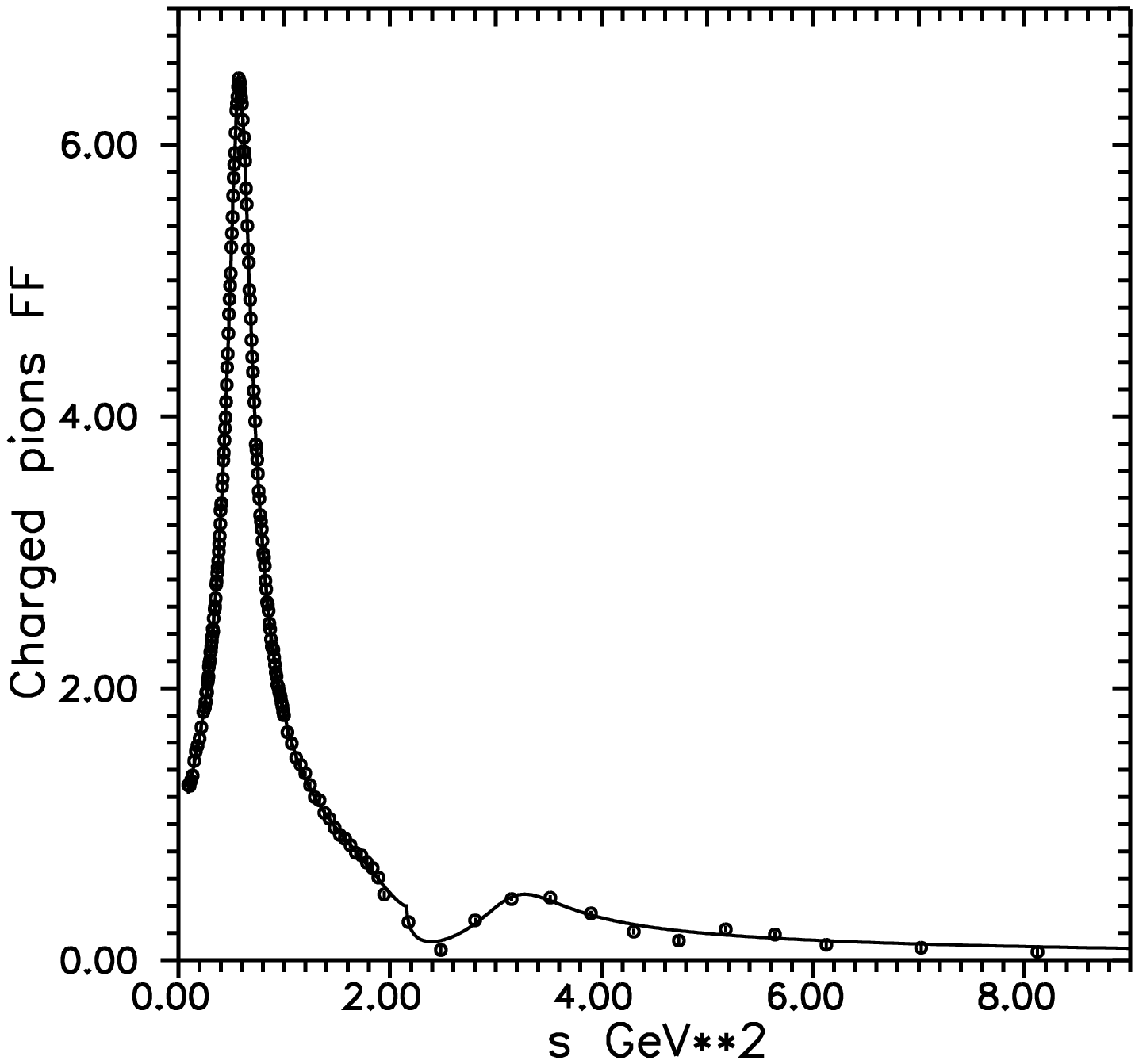}
	\caption{Optimal description of $|F^c_{\pi}(s)|$ data with full line given by Equations (\ref{pioFF})--(\ref{transVmes}) and parameters values of Table~\ref{tab:1}. \label{fig:10}}
\end{figure}

\begin{figure}[ht]
	
	\includegraphics[width=.65\columnwidth]{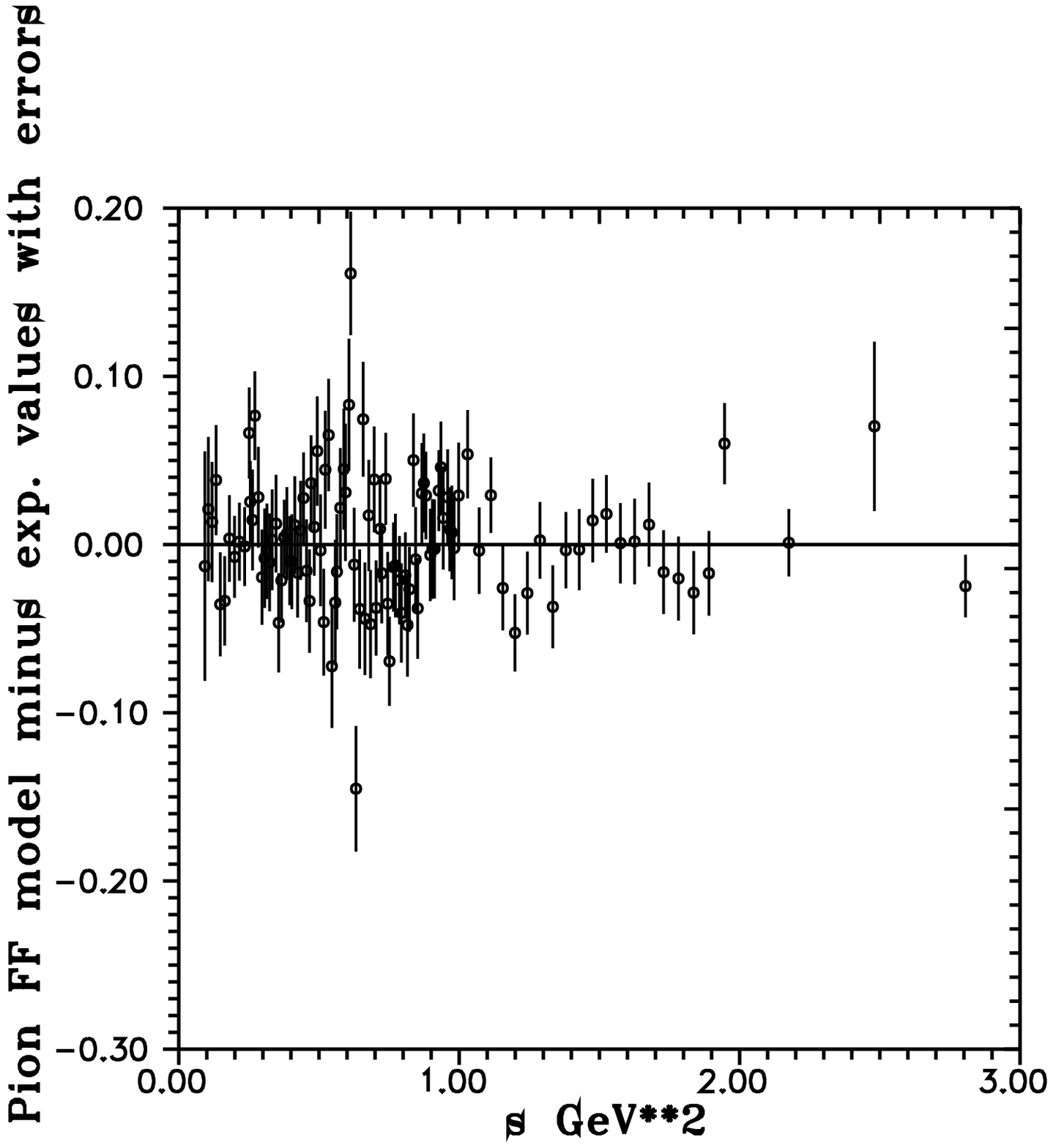}
	\caption{Graphical demonstration of a subtraction of the full line data in Figure~\ref{fig:10} from the pure EM form factor $|F^c_{\pi}(s)|$ data as given in Tables~\ref{fig:2}--\ref{fig:4}	reveal no damped oscillatory~structures. \label{fig:11}}
\end{figure}

A perfect description of the data on $|F^c_{\pi}(s)|$ accumulated in Tables~\ref{fig:2}--\ref{fig:4} \mbox{by Equations~(\ref{pioFF})--(\ref{transVmes})} with the numerical values of parameters of Table~\ref{tab:1} is represented by the full line in Figure~\ref{fig:10}.
If full line data in Figure~\ref{fig:10} are subtracted from the data in Tables~\ref{fig:2}--\ref{fig:4}, no oscillation structures appear as it is demonstrated clearly in Figure~\ref{fig:11}.

\unskip

\section{Conclusions and~Discussions}

The damped oscillating structures from the proton ``effective'' form factor data invite questions about the existence the similar oscillating structures in the case of other strong interacting particles. One can investigate them if solid data on the corresponding form factors exists together with a physically well-founded model, and by means of it the accurate description of these data is achieved. The only suitable candidates for such investigations are the~charged pion and the charged kaon electromagnetic form factor~data.

In this paper, the problem of the pure isovector charged pion electromagnetic form factor data by using the same procedure as in the case of the proton has been investigated. Since oscillation structures have to be concerned of the $\gamma\pi^+\pi^-$ vertex generated by the strong interactions only and the measured $\sigma^{bare}_{tot}(e^+e^-\to \pi^+\pi^-)$ contains some part of $\pi^+\pi^-$ pairs due to electromagnetic isospin violating decay of $\omega(782)\to\pi^+\pi^-$, first an elimination of a contribution of the latter into three the most precise measurements~\cite{Lees3,Xiao,Ablikim7} of $\sigma^{bare}_{tot}(e^+e^-\to \pi^+\pi^- (\gamma))$ with the ISR method has been carried out. With~this aim, first $\sigma^{bare}_{tot}(e^+e^-\to \pi^+\pi^- (\gamma))$ was expressed as the absolute value squared of the sum of the pure charged pion electromagnetic form factor and the $\omega(782)\to\pi^+\pi^-$ contribution expressed by the Breit-Wigner formula (see Equation~(\ref{totcspipi})). Then, by exploiting the identity of the charged pion electromagnetic form factor phase with P-wave isovector $\pi\pi$-scattering phase shift and the most accurate up-to-now data of the latter, the~data on the absolute value of the pure isovector electromagnetic form factor with errors have been obtained by Equation (\ref{gensol}) in Tables~\ref{tab:2}--\ref{tab:4}.
If these data are best possibly described by the adaptation to the charged pion three parametric formula (see dashed lines in Figures~\ref{fig:6} and \ref{fig:8}), damped oscillatory structures appear (Figures~\ref{fig:7} and \ref{fig:9}). If~data are described by well physically founded U\&A model  Equations (\ref{pioFF})--(\ref{transVmes}) (see Figure~\ref{fig:10} with full line) no damped oscillatory structures appear as it is clearly demonstrated in Figure~\ref{fig:11}.

So, the proper usage of the physically well founded theoretical model accurately describing experimental data eliminate the effect of creating the~oscillations.

Further investigations in this sense concerning the proton and also the charged K-meson are in~progress.

\begin{acknowledgments}
This research was funded by the Slovak Grant Agency for Sciences grant number VEGA 2/0105/21.	
\end{acknowledgments}


\begin{thebibliography}{999}
	
\bibitem[Pedlar \em{et~al.}(2005)Pedlar et~al.]{Pedlar}
	Pedlar, T.K.; Cronin-Hennessy, D.; Gao, K.Y.; Gong, D.T.; Hietala, J.; Kubota, Y.; Klein, T.; Lang, B.W.; Li, S.Z.; Poling, R. 
; et al.
	\newblock {Precision measurements of the timelike electromagnetic form factors
		of pion, kaon, and proton}.
	\newblock {\em Phys. Rev. Lett.} {\bf 2005}, {\em 95},~261803.
	\newblock {\url{https://doi.org/10.1103/PhysRevLett.95.261803}}.
	
	
\bibitem[Ablikim \em{et~al.}(2005)Ablikim et~al.]{Ablikim1}
	BES Collaboration; Ablikim, M.; Bai, J.Z.; Ban, Y.; Bian, J.G.; Cai, X.; Chen, H.F.; Chen, H.S.; Chen, H.X.; Chen, J.C.; et al.
	\newblock {Measurement of the cross section for $e^+e^-\to p\bar p$ at
		center-of-mass energies from 2.0 to 3.07 GeV}.
	\newblock {\em Phys. Lett. B} {\bf 2005}, {\em 630},~14--20.
	\newblock {\url{https://doi.org/10.1016/j.physletb.2005.09.044}}.
	
	
\bibitem[Ablikim \em{et~al.}(2016)Ablikim et~al.]{Ablikim2}
	Ablikim, M.; Achasov, M.N.; Ai, X.C.; Albayrak, O.; Albrecht, M.; Ambrose, D.J.; Amoroso, A.; An, F.F.; An, Q.; Bai, J.Z.; et al.
	\newblock {Measurement of the $e^{+} e^{-} \to \pi^{+} \pi^{-}$ cross section
		between 600 and 900 MeV using initial state radiation}.
	\newblock {\em Phys. Lett.} {\bf 2016}, {\em B753},~629--638.
	\newblock {\url{https://doi.org/10.1016/j.physletb.2015.11.043}}.
	
	
\bibitem[Akhmetshin \em{et~al.}(2016)Akhmetshin et~al.]{Akhmetshin1}
	Akhmetshin, R.R.; Amirkhanov, A.N.; Anisenkov, A.V.; Aulchenko, V.M.; Banzarov, V.S.; Bashtovoy, N.S.; Berkaev, D.E.; Bondar, A.E.; Bragin, A.V.; Eidelman, S.I.; et al.
	\newblock Study of the process $e^+e^-\to p\bar p$ in the c.m. energy range
	from threshold to 2 GeV with the CMD-3 detector.
	\newblock {\em Phys. Lett. B} {\bf 2016}, {\em 759},~634--640.
	\newblock {\url{https://doi.org/10.1016/j.physletb.2016.04.048}}.
	
	
\bibitem[Akhmetshin \em{et~al.}(2019)Akhmetshin et~al.]{Akhmetshin2}
	Akhmetshin, R.R.; Amirkhanov, A.N.; Anisenkov, A.V.; Aulchenko, V.M.; Banzarov, V.S.; Bashtovoy, N.S.; Berkaev, D.E.; Bondar, A.E.; Bragin, A.V.; Eidelman, S.I.; et al.
	\newblock Observation of a fine structure in $e^+e^- \to hadrons$ production at
	the nucleon-antinucleon threshold.
	\newblock {\em Phys. Lett. B} {\bf 2019}, {\em 794},~64--68.
	\newblock {\url{https://doi.org/10.1016/j.physletb.2019.05.032}}.
	
	
\bibitem[Ablikim \em{et~al.}(2020)Ablikim et~al.]{Ablikim3}
	Ablikim, M.; Achasov, M.; Adlarson, P.; Ahmed, S.; Albrecht, M.; Alekseev, M.; Amoroso, A.; An, F.;  An, Q.; Anita; et al.
	\newblock Measurement of Proton Electromagnetic Form Factors in $e^+e^-\to
	p\bar p$ in the Energy Region 2.00--3.08 GeV.
	\newblock {\em Phys. Rev. Lett.} {\bf 2020}, {\em 124},~042001.
	\newblock {\url{https://doi.org/10.1103/PhysRevLett.124.042001}}.
	
	
\bibitem[Aubert \em{et~al.}(2006)Aubert et~al.]{Aubert}
Aubert, B.; Barate, R.; Boutigny, D.; Couderc, F.; Karyotakis, Y.; Lees, J.P.; Poireau, V.; Tisserand, V.; Zghiche, A.; Grauges, E.; et al.
	\newblock Study of $e^+e^-\to p\bar p$ using initial state radiation with
	BABAR.
	\newblock {\em Phys. Rev. D} {\bf 2006}, {\em 73},~012005.
	\newblock {\url{https://doi.org/10.1103/PhysRevD.73.012005}}.
	
	
\bibitem[Lees \em{et~al.}(2013{\natexlab{a}})Lees et~al.]{Lees1}
	Lees, J.P.; Poireau, V.; Tisserand, V.; Grauges, E.; Palano, A.; Eigen, G.; Stugu, B.; Brown, D.N.; Kerth, L.T.; Yu, G.; et al.
	\newblock Study of $e^+e^-\to p\bar p$ via initial-state radiation at BABAR.
	\newblock {\em Phys. Rev. D} {\bf 2013}, {\em 87},~092005.
	\newblock {\url{https://doi.org/10.1103/PhysRevD.87.092005}}.
	
	
\bibitem[Lees \em{et~al.}(2013{\natexlab{b}})Lees et~al.]{Lees2}
	Lees, J.P.; Poireau, V.; Tisserand, V.; Grauges, E.; Palano, A.; Eigen, G.; Stugu, B.; Brown, D.N.; Kerth, L.T.;  Yu, G.; et al.
	\newblock {Production of charged pions, kaons, and protons in $e^+e^-$
		annihilations into hadrons at $\sqrt{s}=10.54$ GeV}.
	\newblock {\em Phys. Rev. D} {\bf 2013}, {\em 88},~032011.
	\newblock {\url{https://doi.org/10.1103/PhysRevD.88.032011}}.
	
	
\bibitem[Ablikim \em{et~al.}(2019)Ablikim et~al.]{Ablikim4}
Ablikim, M.; Achasov, M.; Adlarson, P.; Ahmed, S.; Albrecht, M.; Alekseev, M.; Amoroso, A.; An, F.; An, Q.; Bai, Y.; et al.
	\newblock Study of the process $e^+e^-\to p\bar p$ via initial state radiation
	at BESIII.
	\newblock {\em Phys. Rev. D} {\bf 2019}, {\em 99},~092002.
	\newblock {\url{https://doi.org/10.1103/PhysRevD.99.092002}}.
	
	
\bibitem[Ablikim \em{et~al.}(2021)Ablikim et~al.]{Ablikim5}
	Ablikim, M.; Achasov, M.N.; Adlarson, P.; Ahmed, S.; Albrecht, M.; Aliberti, R.; Amoroso, A.; An, M.R.; An, Q.; Bai, X.H.; et al.
	\newblock Measurement of proton electromagnetic form factors in the time-like
	region using initial state radiation at BESIII.
	\newblock {\em Phys. Lett. B} {\bf 2021}, {\em 817},~136328.
	\newblock
	{\url{https://doi.org/https://doi.org/10.1016/j.physletb.2021.136328}}.
	
	
\bibitem[Baldini~Ferroli \em{et~al.}(2012)Baldini~Ferroli, Pacetti, and
	Zallo]{BaPaZa}
	Baldini~Ferroli, R.; Pacetti, S.; Zallo, A.
	\newblock {No Sommerfeld resummation factor in $e^+e^- \to p \bar{p}$?}
	\newblock {\em Eur. Phys. J. A} {\bf 2012}, {\em 48},~33.
	\newblock {\url{https://doi.org/10.1140/epja/i2012-12033-6}}.
	
	
\bibitem[Tomasi-Gustafsson and Rekalo(2001)]{TomRek}
	Tomasi-Gustafsson, E.; Rekalo, M.P.
	\newblock {Search for evidence of asymptotic regime of nucleon electromagnetic
		form-factors from a compared analysis in space- and time-like regions}.
	\newblock {\em Phys. Lett. B} {\bf 2001}, {\em 504},~291--295.
	\newblock {\url{https://doi.org/10.1016/S0370-2693(01)00312-4}}.
	
	
\bibitem[Dunning \em{et~al.}(1966)Dunning, Chen, Cone, Hartwig, Ramsey, Walker,
	and Wilson]{DChCHRWW}
	Dunning, J.R.; Chen, K.W.; Cone, A.A.; Hartwig, G.; Ramsey, N.F.; Walker, J.K.;
	Wilson, R.
	\newblock Quasi-Elastic Electron-Deuteron Scattering and Neutron Form Factors.
	\newblock {\em Phys. Rev.} {\bf 1966}, {\em 141},~1286--1297.
	\newblock {\url{https://doi.org/10.1103/PhysRev.141.1286}}.
	
	
\bibitem[Akhiezer and Rekalo(1968)]{AkhRek1}
	Akhiezer, A.I.; Rekalo, M.P.
	\newblock {Polarization phenomena in electron scattering by protons in the high
		energy region}.
	\newblock {\em Sov. Phys. Dokl.} {\bf 1968}, {\em 13},~572.
	
	
\bibitem[Akhiezer and Rekalo(1974)]{AkhRek2}
	Akhiezer, A.I.; Rekalo, M.P.
	\newblock {Polarization effects in the scattering of leptons by hadrons}.
	\newblock {\em Sov. J. Part. Nucl.} {\bf 1974}, {\em 4},~277.
	
	
\bibitem[Jones \em{et~al.}(2000)Jones et~al.]{Jones}
	Jones, M.K.; Aniol, K.A.; Baker, F.T.; Berthot, J.; Bertin, P.Y.; Bertozzi, W.; Besson, A.; Bimbot, L.; Boeglin, W.U.; Brash, E.J.; et al.
	\newblock {${G}_{{E}_{p}}/{G}_{{M}_{p}}$ Ratio by Polarization Transfer in
		$\stackrel{\ensuremath{\rightarrow}}{e}\mathit{p}\phantom{\rule{0ex}{0ex}}\ensuremath{\rightarrow}\phantom{\rule{0ex}{0ex}}\mathit{e}\stackrel{\ensuremath{\rightarrow}}{p}$}.
	\newblock {\em Phys. Rev. Lett.} {\bf 2000}, {\em 84},~1398--1402.
	\newblock {\url{https://doi.org/10.1103/PhysRevLett.84.1398}}.
	
	
\bibitem[Gayou \em{et~al.}(2002)Gayou et~al.]{Gayou}
	Gayou, O.; Aniol, K.A.; Averett, T.; Benmokhtar, F.; Bertozzi, W.; Bimbot, L.; Brash, E.J.; Calarco, J.R.; Cavata, C.; Chai, Z.; et al.
	\newblock {Measurement of ${G}_{{E}_{p}}/{G}_{{M}_{p}}$ in
		$\stackrel{\ensuremath{\rightarrow}}{e}\mathit{p}\ensuremath{\rightarrow}\mathit{e}\stackrel{\ensuremath{\rightarrow}}{p}$
		to
		${\mathit{Q}}^{2}\phantom{\rule{0ex}{0ex}}=\phantom{\rule{0ex}{0ex}}5.6{\mathrm{GeV}}^{2}$}.
	\newblock {\em Phys. Rev. Lett.} {\bf 2002}, {\em 88},~092301.
	\newblock {\url{https://doi.org/10.1103/PhysRevLett.88.092301}}.
	
	
\bibitem[Punjabi \em{et~al.}(2005)Punjabi et~al.]{Punjabi}
	Punjabi, V.; Perdrisat, C.F.; Aniol, K.A.; Baker, F.T.; Berthot, J.; Bertin, P.Y.; Bertozzi, W.; Besson, A.; Bimbot, L.; Boeglin, W.U.; et al.
	\newblock {Proton elastic form factor ratios to
		${Q}^{2}=3.5\phantom{\rule{0.3em}{0ex}}\text{GeV}{}^{2}$ by polarization
		transfer}.
	\newblock {\em Phys. Rev. C} {\bf 2005}, {\em 71},~055202.
	\newblock {\url{https://doi.org/10.1103/PhysRevC.71.055202}}.
	
	
\bibitem[Zhan \em{et~al.}(2011)Zhan et~al.]{Zhan}
	Zhan, X.; Allada, K.; Armstrong, D.S.; Arrington, J.; Bertozzi, W.; Boeglin, W.; Chen, J.-P.; Chirapatpimol, K.; Choi, S.; Chudakov, E.; et al.
	\newblock {High-precision measurement of the proton elastic form factor ratio
		$\mu_{p}G_{E}/G_{M}$ at low $Q^2$}.
	\newblock {\em Phys. Lett. B} {\bf 2011}, {\em 705},~59--64.
	\newblock
	{\url{https://doi.org/https://doi.org/10.1016/j.physletb.2011.10.002}}.
	
	
\bibitem[Puckett and others.(2017)]{Puckett2}
	Puckett, A.J.R.; Brash, E.J.; Jones, M.K.; Luo, W.; Meziane, M.; Pentchev, L.; Perdrisat, C.F.; Punjabi, V.; Wesselmann, F.R.; Afanasev, A.; et al.
	\newblock {Polarization transfer observables in elastic electron-proton
		scattering at ${Q}^{2}=2.5$, 5.2, 6.8, and $8.5 {\mathrm{GeV}}^{2}$}.
	\newblock {\em Phys. Rev. C} {\bf 2017}, {\em 96},~055203.
	\newblock {\url{https://doi.org/10.1103/PhysRevC.96.055203}}.
	
	
\bibitem[Bianconi and Tomasi-Gustafsson(2015)]{BianTom}
	Bianconi, A.; Tomasi-Gustafsson, E.
	\newblock {Periodic interference structures in the timelike proton form
		factor}.
	\newblock {\em Phys. Rev. Lett.} {\bf 2015}, {\em 114},~232301.
	\newblock {\url{https://doi.org/10.1103/PhysRevLett.114.232301}}.
	
	
\bibitem[Ablikim \em{et~al.}(2021)Ablikim et~al.]{Ablikim6}
	Ablikim, M.
	\newblock {Oscillating features in the electromagnetic structure of the
		neutron}.
	\newblock {\em Nat. Phys.} {\bf 2021}, {\em 17},~1200--1204.
	\newblock {\url{https://doi.org/10.1038/s41567-021-01345-6}}.
	
	
\bibitem[Bianconi and Tomasi-Gustafsson(2016)]{BianTom2}
	Bianconi, A.; Tomasi-Gustafsson, E.
	\newblock Phenomenological analysis of near-threshold periodic modulations of
	the proton timelike form factor.
	\newblock {\em Phys. Rev. C} {\bf 2016}, {\em 93},~035201.
	\newblock {\url{https://doi.org/10.1103/PhysRevC.93.035201}}.
	
	
\bibitem[Tomasi-Gustafsson and Pacetti(2022)]{BianTom3}
	Tomasi-Gustafsson, E.; Pacetti, S.
	\newblock Interpretation of recent form factor data in terms of an advanced
	representation of baryons in space and time.
	\newblock {\em Phys. Rev. C} {\bf 2022}, {\em 106},~035203.
	\newblock {\url{https://doi.org/10.1103/PhysRevC.106.035203}}.
	
	
\bibitem[Lorenz \em{et~al.}(2015)Lorenz, Hammer, and Mei\ss{}ner]{Lorenz}
	Lorenz, I.T.; Hammer, H.W.; Mei\ss{}ner, U.G.
	\newblock {New structures in the proton-antiproton system}.
	\newblock {\em Phys. Rev. D} {\bf 2015}, {\em 92},~034018.
	\newblock {\url{https://doi.org/10.1103/PhysRevD.92.034018}}.
	
	
\bibitem[Lees \em{et~al.}(2012)Lees et~al.]{Lees3}
	Lees, J.P.; Poireau, V.; Tisserand, V.; Tico, J.G.; Grauges, E.; Palano, A.; Eigen, G.; Stugu, B.; Brown, D.N.; Kerth, L.T.; et al.
	\newblock {Precise Measurement of the $e^+ e^- \to \pi^+\pi^- (\gamma)$ Cross
		Section with the Initial-State Radiation Method at BABAR}.
	\newblock {\em Phys. Rev.} {\bf 2012}, {\em D86},~032013.
	\newblock {\url{https://doi.org/10.1103/PhysRevD.86.032013}}.
	
	
\bibitem[Xiao \em{et~al.}(2018)Xiao, Dobbs, Tomaradze, Seth, and
	Bonvicini]{Xiao}
	Xiao, T.; Dobbs, S.; Tomaradze, A.; Seth, K.K.; Bonvicini, G.
	\newblock Precision measurement of the hadronic contribution to the muon
	anomalous magnetic moment.
	\newblock {\em Phys. Rev. D} {\bf 2018}, {\em 97},~032012.
	\newblock {\url{https://doi.org/10.1103/PhysRevD.97.032012}}.
	
	
\bibitem[Ablikim \em{et~al.}(2021)Ablikim et~al.]{Ablikim7}
	Ablikim, M.; Achasov, M.N.; Adlarson, P.; Ahmed, S.; Albrecht, M.; Aliberti, R.; Amoroso, A.; An, Q.; Bai, X.H.; Bai, Y.; et al.
	\newblock Corrigendum to ``Measurement of the $e^+e^-\to \pi^+ \pi^-$ cross
	section between 600 and 900 MeV using initial state radiation'' [Phys. Lett.
	B 753 (2016) 629–638].
	\newblock {\em Phys. Lett. B} {\bf 2021}, {\em 812},~135982.
	\newblock
	{\url{https://doi.org/https://doi.org/10.1016/j.physletb.2020.135982}}.
	
	
\bibitem[Dubnicka and Dubnickova(2010)]{DD}
	Dubnicka, S.; Dubnickova, A.
	\newblock {Analyticity in a phenomenology of electro-weak structure of
		hadrons}.
	\newblock {\em Acta Phys. Slov.} {\bf 2010}, {\em 60},~1--153.
	
	
\bibitem[Zyla \em{et~al.}(2020)Zyla et~al.]{PDG}
	Particle Data Group; Zyla, P.A.; Barnett, R.M.; Beringer, J.; Dahl, O.; Dwyer, D.A.; Groom, D.E.; Lin, C.-J.; Lugovsky, K.S.; Pianori, E.; et al.
	\newblock {Review of Particle Physics}.
	\newblock {\em PTEP} {\bf 2020}, {\em 2020},~083C01.
	\newblock {\url{https://doi.org/10.1093/ptep/ptaa104}}.
	
	
\bibitem[Biagini \em{et~al.}(1991)Biagini, Dubnicka, Etim, and Kolar]{BDEK}
	Biagini, M.E.; Dubnicka, S.; Etim, E.; Kolar, P.
	\newblock {Phenomenological evidence for a third radial excitation of rho
		(770)}.
	\newblock {\em  Il Nuovo Cimento D } {\bf 1991}, {\em 104},~363--370.
	\newblock {\url{https://doi.org/10.1007/BF02799144}}.
	
	
\bibitem[Garcia-Martin \em{et~al.}(2011)Garcia-Martin, Kaminski, Pelaez,
	Ruiz~de Elvira, and Yndurain]{Garcia}
	Garcia-Martin, R.; Kaminski, R.; Pelaez, J.R.; Ruiz~de Elvira, J.; Yndurain,
	F.J.
	\newblock {The Pion-pion scattering amplitude. IV: Improved analysis with once
		subtracted Roy-like equations up to 1100 MeV}.
	\newblock {\em Phys. Rev.} {\bf 2011}, {\em D83},~074004.
	\newblock {\url{https://doi.org/10.1103/PhysRevD.83.074004}}.
	
	
\bibitem[Bartoš \em{et~al.}(2017)Bartoš, Dubnička, Liptaj, Dubničková, and
	Kamiński]{Bartos}
	Bartoš, E.; Dubnička, S.; Liptaj, A.; Dubničková, A.Z.; Kamiński, R.
	\newblock {What are the correct $\rho^0$(770) meson mass and width values?}
	\newblock {\em Phys. Rev.} {\bf 2017}, {\em D96},~113004.
	\newblock {\url{https://doi.org/10.1103/PhysRevD.96.113004}}.
	
\end{thebibliography}


\end{document}